\documentclass[final]{aa}

\usepackage{graphicx}
\usepackage{tikz}     
\usepackage{txfonts}

\usepackage{amsmath}	
\usepackage{subfig}
\usepackage{float}
\usepackage{xcolor,soul}
\usepackage{xspace}
\usepackage{amssymb}
\usepackage{siunitx}
\usepackage{fontawesome5}
\usepackage{hyperref}

\DeclareMathOperator*{\argmax}{arg\,max}
\DeclareMathOperator*{\argmin}{arg\,min}
\DeclareMathOperator{\E}{\mathbb{E}}
\DeclareMathOperator{\D}{\mathbb{D}}
\DeclareMathOperator{\X}{\Tilde{X}}
\DeclareMathOperator{\Z}{\Tilde{Z}}
\DeclareMathOperator{\x}{\Tilde{x}}
\DeclareMathOperator{\z}{\Tilde{z}}
\DeclareMathOperator{\fieldX}{\mathbb{X}}
\DeclareMathOperator{\Lagr}{\mathbb{L}}
\DeclareMathOperator{\ssim}{\mathbb{S}}

\newcommand{\IsolatedSims}[0]{\texttt{IsolatedSims}\xspace}
\newcommand{\BlendedSims}[0]{\texttt{BlendedSims}\xspace}

\newcommand{\TestFields}[0]{\texttt{TestFields}\xspace}
\newcommand{\MADNESS}[0]{\texttt{MADNESS}\@\xspace}
\newcommand{\VAEdeblender}[0]{\texttt{VAE-deblender}\xspace}
\newcommand{\scarlet}[0]{\textsc{scarlet}\@\xspace}
\newcommand{\SurveyCodex}[0]{\textsc{SurveyCodex}\xspace}
\newcommand{\btk}[0]{\texttt{BTK}\xspace}
\newcommand{\TensorFlow}[0]{\textsc{TensorFlow}\xspace}
\newcommand{\TFP}[0]{\textsc{TensorFlow-Probability}\xspace}
\newcommand{\galsim}[0]{\textsc{GalSim}\xspace}
\newcommand{\SourceExtractor}[0]{\textsc{SourceExtractor}\xspace}
\newcommand{\BlendingToolkit}[0]{\textsc{Blending-Toolkit}\xspace}
\newcommand{\CatSim}[0]{CatSim\xspace}

\makeatletter
\renewcommand*\aa@pageof{, page \thepage{} of \pageref*{LastPage}}
\makeatother

\begin{document} 
\nolinenumbers

   \title{\MADNESS deblender}

   \subtitle{Maximum A posteriori with Deep NEural networks for Source Separation}

   \author{B. Biswas
          \inst{1}\thanks{\email{biswas@apc.in2p3.fr}}
          \and
          E. Aubourg
          \inst{1}
          \and
          A. Boucaud
          \inst{1}
          \and
          A. Guinot
          \inst{1,2}
          \and
          J. Lao
          \inst{3}
          \and
          C. Roucelle
          \inst{1}
          \and 
          {the LSST Dark Energy Science Collaboration}
          }

   \institute{
    Université Paris Cité, CNRS, AstroParticule et Cosmologie, F-75013, Paris, France
    \and
    McWilliams Center for Cosmology, Department of Physics, Carnegie Mellon University, Pittsburgh, PA 15213, USA
    \and
    Google Research, Google, Gustav-Gull-Platz 1, 8004, Zurich, Switzerland
}

   \date{14 August 2024 / Accepted 7 May 2025}

  \abstract
   {Due to the unprecedented depth of the upcoming ground-based Legacy Survey of Space and Time (LSST) at the Vera C. Rubin Observatory, approximately two-thirds of the galaxies are likely to be affected by blending - the overlap of physically separated galaxies in images. 
   Thus, extracting reliable shapes and photometry from individual objects will be limited by our ability to correct blending and control any residual systematic effect. Deblending algorithms tackle this issue by reconstructing the isolated components from a blended scene, but the most commonly used algorithms often fail to model complex, realistic galaxy morphologies.}
   {As part of an effort to address this major challenge, we present \MADNESS, which takes a data-driven approach and combines pixel-level multi-band information to learn complex priors for obtaining the maximum a posteriori solution of deblending.}
   {\MADNESS is based on deep neural network architectures, namely variational auto-encoders and normalizing flows. The variational auto-encoder reduces the high-dimensional pixel space into a lower-dimensional space, while the normalizing flow models a data-driven prior in this latent space. Together, these neural networks enable one to obtain the maximum a posteriori solution in the latent space.}
   {Using a simulated test dataset with galaxy models for a 10-year LSST survey and a galaxy density ranging from $48$ to $80$ galaxies per arcmin\textsuperscript{2}, we characterized the aperture-photometry, \textit{g-r} color, structural similarity index, and pixel cosine similarity of the galaxies reconstructed by \MADNESS. 
   We compared our results against state-of-the-art deblenders including \scarlet.
   With the \textit{r}-band of LSST as an example, we show that \MADNESS performs better than \scarlet in all the metrics. 
  For instance, the average absolute value of relative flux residual in the \textit{r}-band for \MADNESS is approximately $29\%$ lower than that of \scarlet.
  The code is publicly available on GitHub \href{https://github.com/b-biswas/MADNESS}{\faGithub}.}
   {}


   \maketitle

\nolinenumbers
\section{Introduction}

One of the main goals of the next-generation cosmological surveys such as the Legacy Survey of Space and Time \cite[LSST\footnote{\url{https://www.lsst.org/}},][]{2009arXiv0912.0201L} at the Vera C. Rubin Observatory, ESA's Euclid\footnote{\url{https://www.esa.int/Science_Exploration/Space_Science/Euclid}} telescope \citep{2010arXiv1001.0061R}, and the Nancy Grace Roman Space Telescope\footnote{\url{https://roman.gsfc.nasa.gov/}} \citep{2013arXiv1305.5422S}, is to understand the fundamental nature of dark energy.
Among the cosmological probes that can be used for these studies, the use of cosmic shear, in particular, is expected to make a major step forward with these stage IV surveys. 
Cosmic shear corresponds to the systematic distortions in the shapes of astrophysical objects due to the bending of the path of light by inhomogeneous matter distribution present between us and the distant source galaxies (so-called gravitational lensing). 
Studying these correlated distortions in tomographic redshift bins allows one to map the underlying dark matter distribution and its evolution in time, and thus constrain dark energy parameters.
However, in the weak regime of gravitational lensing, these distortions imprinted on galaxies are extremely small, of the order of a few percent. 
This requires high statistical precision to impose strong constraints on dark energy and forces the community to probe the universe to an unprecedented depth with these wide-field surveys.

With increased statistical precision, new challenges in data analysis are also expected.
One such challenge is the blending of galaxies, which occurs when physically separated objects lie along the same line of sight and overlap in the pixels. 
The increase in object density due to the greater depth of observations will make the effect of blending significant in upcoming surveys.
In the case of ground-based surveys, such as the LSST, the point-spread-function (PSF) contribution from the Earth's atmosphere will lead to an additional blurring of images and increase the effect of blending. 
In the wide survey of Hyper Suprime-Cam \cite[HSC\footnote{\url{https://hsc.mtk.nao.ac.jp/ssp/}},][]{10.1093/pasj/psx066}, with a limiting \textit{i}-band magnitude $\approx 26$, around $58\%$ of the objects were observed to be blended \citep{10.1093/pasj/psx080}. 
The problem is expected to be more severe in the case of LSST since it aims to reach $i\approx 27$ after 10 years of operation.
As a consequence, it is expected that around $62\%$ of LSST objects will have at least a $1\%$ flux contribution from neighboring sources \cite[see][]{2021JCAP...07..043S}.

The extent to which the systematic errors can be controlled will determine how well cosmological models can be constrained \cite[see ][for a detailed discussion on weak lensing systematic effects]{doi:10.1146/annurev-astro-081817-051928}, and blending is going to be a major part of these systematic errors.
For example, blending can bias photometric redshift (photo-\textit{z}) measurements that will in turn affect the downstream analysis of most cosmological probes, including weak lensing.
To measure galaxy properties on blended scenes, the effects of blending must be mitigated before processing the data with pipelines that are designed to operate with single sources. The inverse problem of reconstructing the individual isolated components from a blended scene is called deblending.
Another approach for correcting weak lensing measurements is to calibrate for the effect of blending, among other systematic effects, using image simulations \citep{10.1093/mnras/stab2870, Sheldon_2020}. However, the photometry of the sources is still affected, and it remains to be demonstrated that these simulations would be sufficient to capture the entirety of the blending impacts for LSST weak lensing analysis. 

To make matters worse, a considerable fraction of these blends are likely to go undetected. 
These cases in which the detection algorithm fails to identify multiple sources among the blended galaxies are called \lq\lq unrecognized blends\rq\rq.
Typically, this can happen when the centers of two objects are closer to each other than a fraction of the PSF or when one of the sources is much brighter than the other, so that the detection algorithm cannot distinguish the fainter source.
Although it is more optimal to perform joint detection and deblending, combining the two steps can be computationally very expensive.
Most deblenders assume prior detection of objects; therefore, they rely on robust detection algorithms to deblend galaxies from a blended scene efficiently.
Synergies between different surveys can also mitigate the effects of unrecognized blends. 
Since space-based surveys do not have an atmospheric contribution to the PSF, the sources are more resolved and can help provide priors for detections in ground-based surveys, where the blending effect is stronger due to the atmosphere \cite[see][]{osti_1833317}. 
In fact, \cite{10.1093/mnras/staa3062} showed that joint pixel-level processing of ground- and space-based data can help improve the quality of deblended models, motivating further development of algorithms that are capable of simultaneously processing data from multiple surveys. 

One of the most commonly used deblending algorithms, \texttt{SExtractor} \cite[developed by][]{1996A&AS..117..393B},
performs image segmentation by looking for connected components at various thresholds and computes the deblending solution independently on each band or uses a dual mode to use one band as the reference (e.g., the \textit{r}-band) to obtain photometry in another. 
Such a method cannot realistically deblend a galaxy because it will assign pixels in an overlapping case to one of the galaxies without trying to redistribute the fluxes.
This results in strong limitations in terms of both the morphology and the fluxes of the reconstructions.

Although the Sloan Digital Sky Survey (SDSS) deblender by \cite{Lupton2005SDSSIP} took a step further and assumed symmetry properties of galaxies as prior, it still performed the deblending individually in different bands.
Other approaches, such as Lambda Adaptive Multi-Band Deblending Algorithm in R (\texttt{LAMBDAR}; \citealp{10.1093/mnras/stw832})
use higher-resolution aperture priors to obtain consistent matched photometry across a range of photometric data with various resolutions.
Another commonly used algorithm, called TheTractor \citep{2016ascl.soft04008L}, optimizes astronomical object models to find the optimal model that maximizes the likelihood.
More recently, \cite{MELCHIOR2018129} have developed the \scarlet deblender, which combines information from different bands and uses a constrained matrix factorization to separate overlapping sources.
\scarlet uses proximal gradient descent to impose constraints such as the symmetry and monotonicity of light profiles on the reconstructed components.
The \scarlet deblender makes it possible to incorporate physical information such as noise and the PSF into the algorithm so that knowledge about the data can be leveraged.
Although such an approach is capable of handling more complex blended scenes and provides a significant improvement over previous deblenders, modeling complex morphologies is still a challenge because of the strong regularizations. 

The use of deep learning is gaining ground in the interest of predicting more complex galaxy shapes.
In an approach very similar to \scarlet and also closely related to our work, \cite{https://doi.org/10.48550/arxiv.1912.03980} replaced the regularizations by modeling a data-driven prior using differentiable deep generative models with explicit likelihoods.
This approach solves the maximum a posteriori (MAP) solution of the inverse problem by using autoregressive models such as the \texttt{PixelCNN++} \citep{salimans2017pixelcnn} to compute a likelihood based on a pixel-level galaxy prior. 
Although such a deep learning architecture is quite interpretable, the pixel-wise auto-regressive nature negatively impacts the speed, which is a major concern for upcoming surveys such as the LSST. 

Several other deep learning architectures based on generative models have already shown promising deblending results. 
\cite{10.1093/mnras/stz575} used a branched generative adversarial network (GAN) to deblend two galaxies from a stamp. 
\cite{10.1093/mnras/staa3062} used a modified variational autoencoder (VAE) architecture in which the encoder is trained to remove non-central galaxies and project only the central galaxy into the latent space; the decoder is trained to map from the latent space representation of an isolated galaxy to its image.
Using a similar VAE model as one of its components, \cite{hansen2022scalable} proposed Bayesian Light Source Separator (BLISS), which is a scalable approach to perform Bayesian detection and deblending using a combination of three convolutional encoder networks responsible for detection, star-galaxy classification, and morphology estimation.
Unlike other approaches, BLISS aims to produce a probabilistic catalog of galaxies that can be used by downstream scientific analysis.
More recently, \citet{sampson2024scorematching} introduced \textsc{scarlet}\oldstylenums{2}, which uses neutral score matching to obtain the priors for \scarlet in order to predict more complex morphologies.
Going beyond generative models, other deep learning techniques have also been shown to be effective for deblending. For example, \citet{PhysRevD.106.063023} used a residual dense neural network to extract galaxies iteratively from a field until no galaxies were found. 

Deep learning methods using convolutional neural networks provide an easy way to combine multichannel and multi-instrument information \citep{10.1093/mnras/staa3062}. Although architectures that deblend with a single forward pass are extremely fast at obtaining results, they lack explicit optimization over the total flux. 
To find a middle ground between the VAE deblender proposed by \citet{10.1093/mnras/staa3062} and MAP estimate obtained by \citet{MELCHIOR2018129} and \citet{https://doi.org/10.48550/arxiv.1912.03980}, we have developed a deblender called \MADNESS\footnote{\url{https://github.com/LSSTDESC/madness}, release \texttt{v1.0.0}}, which stands for Maximum A posteriori solution with Deep NEural networks for Source Separation.
Although our work can be extended to obtain a probabilistic catalog and potentially be combined with the detection and classification steps of BLISS, the focus for now is on obtaining the MAP solution.
Similar to most deblenders described above, \MADNESS assumes that the central pixels of galaxies in the field are known and the goal is to reconstruct the isolated components.
As an extension to the deblender proposed by \cite{10.1093/mnras/staa3062}, we coupled the VAE with a normalizing flow introduced by \cite{https://doi.org/10.48550/arxiv.1505.05770}.
The combination of the two neural networks is used in a way that the MAP estimate can be obtained by a gradient descent in the latent space.
By doing so, we show that deep learning can be used efficiently to perform an optimization rather than a single pass through a feed-forward network.

We first describe the simulations used for training and testing \MADNESS in Section \ref{sec:Dataset}. 
In Section \ref{sec:Method}, we dive deeper into variational autoencoders and normalizing flows and discuss in detail the architecture, training of the models, and the methodology of our deblender.
We present the results of our algorithm and compare it with state-of-the-art deblenders in Section \ref{sec:Results} using various metrics.
Finally, in Section \ref{sec: towards real data} we discuss potential next steps to prepare our algorithm for processing real data, before concluding in Section \ref{sec:Conclusions}.

\section{Dataset}
\label{sec:Dataset}

\begin{figure}
\centering
\includegraphics[width=.49 \textwidth]{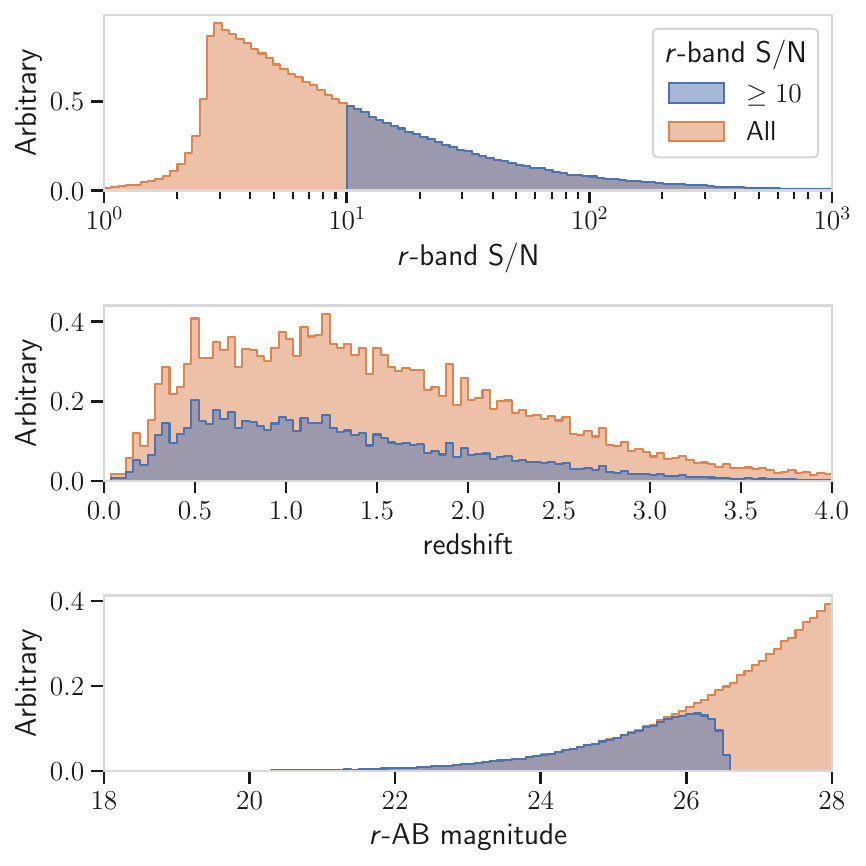}
\caption{Distribution of galaxies from the \CatSim catalog before (orange) and after (blue) the S/N cut for \textit{r}-band S/N (see Appendix \ref{section: S/N computation}), redshift (from catalog), and \textit{r}-AB magnitude (from catalog).}
\label{fig: catalog distribution}
\end{figure}

\begin{figure}[t]
     \centering
     \subfloat[]{
         \centering
         \includegraphics[width=.49\textwidth]{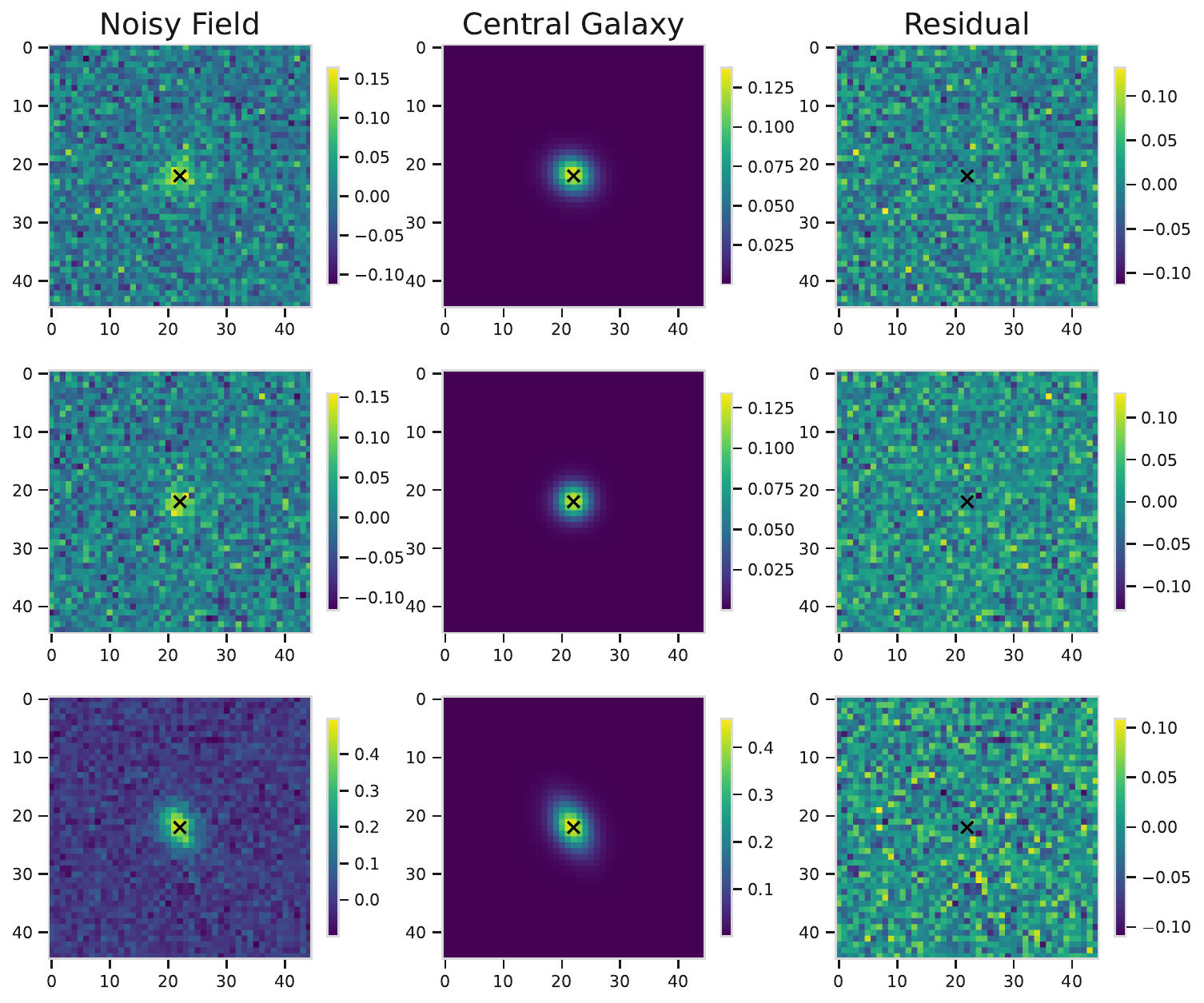}
         \label{fig: isolated examples}
     }
     \hfill
     \subfloat[]{
         \centering
         \includegraphics[width=.49\textwidth]{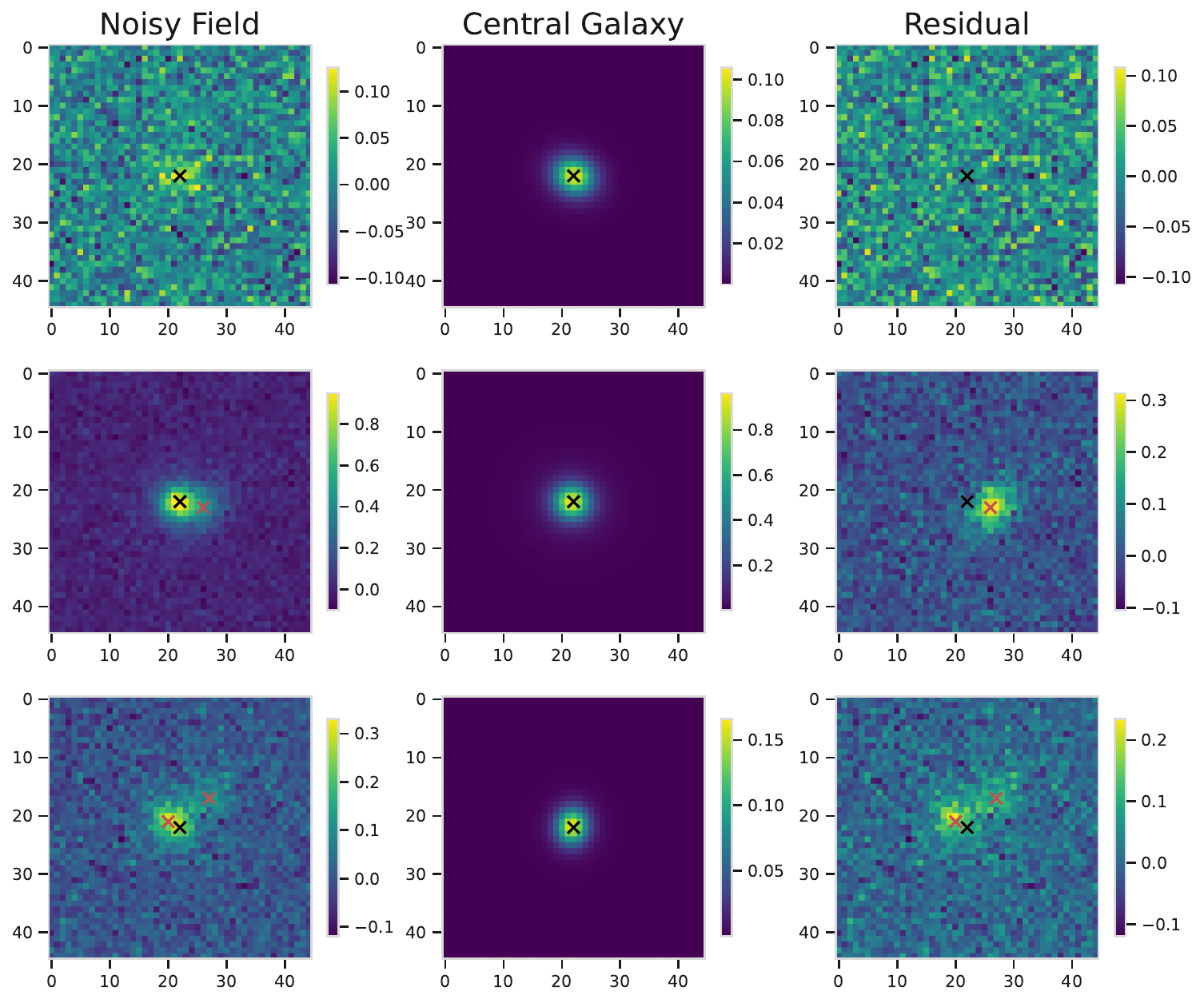}
         \label{fig: blended examples}
     }

    \caption{Examples of normalized simulations in the \textit{r}-band, used for training and validating the variational autoencoder and the normalizing flow.
    Panel (a) shows galaxies from the \IsolatedSims dataset, while panel (b) shows galaxies from the \BlendedSims dataset.
    The central galaxy in the stamps is marked in black while the neighbors in the \BlendedSims are marked in red. Each row represents the same scene, and the first column shows the noisy field; the second column shows the isolated noiseless central galaxy; the third column shows the residual field after subtracting the central galaxy from the noisy field. The color bar for each plot shows the normalized electron counts in the pixels.}
    \label{fig: simulation examples}
\end{figure}

To train and test our deblender, we use galaxies from the \CatSim\footnote{\url{https://www.lsst.org/scientists/simulations/catsim}} galaxy catalog that has been prepared for the LSST catalog simulator \citep{2014SPIE.9150E..14C}. 
The galaxies in the catalog are defined as a sum of bulge and disk models covering a redshift range of $0<z<6$ with a realistic distribution of shapes and sizes.
For each galaxy, all parameters related to shape and size are assumed to be the same in all bands, but not the magnitudes.
For this study, we used a one-square-degree subset that has been previously used by the community to understand the effects of blending on cosmic shear estimation \citep{2021JCAP...07..043S}.
This one square degree of data consists of $\approx 858k$ galaxies with a magnitude up to $28$ in the \textit{r}-filter. 
We further imposed a signal-to-noise ratio (S/N) cut of 10 on the \textit{r}-filter using the S/N definition described in Appendix \ref{section: S/N computation}. 
Figure \ref{fig: catalog distribution} shows the distribution of redshift and \textit{r}-AB magnitude of the galaxies before and after the S/N cut.
Finally, we are left with roughly $317k$ galaxies of which we keep \num{200000} for training and validation and the remaining for testing.

The galaxies are simulated under $10$-year LSST conditions with two $15-$second exposures, each per visit, a gain of $1$, standard zero-point, and sky background brightness computed using the expected number of visits in each filter of LSST \citep{Ivezic_2019}. 
Besides the sky background, we also added Poisson noise due to the flux in each pixel.
We used the \texttt{Kolmogorov} model for the PSF, which is held constant over all the generated images but is different for each band; the width of the PSF is taken from \citet{Ivezic_2019}, assuming an airmass of  $1.2$.
These conditions for the simulations are fetched from \SurveyCodex\footnote{\url{https://github.com/LSSTDESC/surveycodex}} Version 1.2.0
 and summarized in Table \ref{tab:simulations}.

\begin{table}
 \caption{Specifications used for each filter in simulating galaxies from the \CatSim galaxy catalog for training and testing the deblender.}
 \label{tab:simulations}
 \begin{tabular}{c| c | c | c | c}
  \hline
  Filter & zero point & sky background & PSF FWHM & visits\\
  \hline
  u & 26.40 & 22.99 & 0.90 & 56\\[2pt]
  g & 28.26 & 22.26 & 0.86 & 80\\[2pt] 
  r & 28.10 & 21.20 & 0.81 & 184\\[2pt] 
  i & 27.78 & 20.48 & 0.79 & 184\\[2pt] 
  y & 27.39 & 19.60 & 0.77 & 160\\[2pt] 
  z & 26.56 & 18.61 & 0.76 & 160\\[2pt] 
  
  \hline
 \end{tabular}
\end{table}

\subsection{Training and validation data}

We used the \BlendingToolkit\footnote{\url{https://github.com/LSSTDESC/BlendingToolKit/}} (\btk, \citealp{Mendoza_2024BTK}) to generate postage stamps for LSST as it allows easy control over the number and location of galaxies in the stamps.
\btk uses \galsim\footnote{\url{https://github.com/GalSim-developers/GalSim}} \citep{galsim2015} internally to simulate the galaxies. 
Using \btk we generated stamps of size $45 \times 45$ pixels in each band, which corresponds to 9 \arcsec$\times$ 9 \arcsec, and the center of the galaxy to be deblended is always located within the central pixel of the stamp.

Of the \num{200000} galaxies kept for training and validation, we used $75\%$ for training and the remaining $25\%$ for validation. We generated two distinct datasets:
\IsolatedSims, which contains only one galaxy located at the center of each stamp; \BlendedSims, consisting of stamps that are extracted from a field with up to 3 galaxies, the centers of which lie within a box of $20\times 20$ pixels.
With each galaxy in the fields as a center, a stamp of $45 \times 45$ pixels is extracted.

The purpose of the \BlendedSims dataset is to retrain the encoder so that it can initialize the latent space for faster convergence of the MAP optimization, as described in Section \ref{subsection: deblender}. 
The exact distribution of the number of galaxies in the blends will not have a significant impact on the results of \MADNESS. 
So we let the number of galaxies in the fields used for extracting \BlendedSims vary uniformly between 1 and 3. In Figure \ref{fig: simulation examples}, we show sample images used to train our network. 
  
\subsection{Test data}
\label{subsection: Test Data}

The test dataset, hereafter \TestFields, was also generated using \btk 
but excluding the galaxies that were used for training and validation.
Unlike the \IsolatedSims and \BlendedSims datasets; which contain small postage stamps, here we simulate multiple galaxies in random positions over a much larger field and test the results of our deblending algorithm using the true centers of the galaxies.
We simulate fields of size $205 \times 205$ pixels with the centers of galaxies placed randomly only in the inner $150 \times 150$ pixels ($0.25$ arcmin\textsuperscript{2}) to ensure that all pixels of the predicted galaxies lie within the field.  
We used rejection sampling to ensure that the centers of galaxies do not lie within the same pixel.
The \TestFields simulations contain $12$ to $20$ galaxies in each field, corresponding to  $48$ to $80$ galaxies per arcmin\textsuperscript{2}. 
The density of galaxies is higher than the typical values we expect for LSST so that the deblenders can be evaluated and compared to each other for a higher blending rate; tests repeated with a more LSST-like density showed consistent results.
In total, we generate \num{6000} such fields with a total of \num{95972} galaxies.

\section{Method}
\label{sec:Method}

\MADNESS uses a combination of two neural network architectures: a VAE and a normalizing flow.
The goal of the VAE is to learn low-dimension latent-space representations of galaxies so that it can be used as a generative model, while the normalizing flow is used to model the underlying aggregate posterior distribution of galaxies in the latent space. 
Once trained, we can sample from the distribution learned by the normalizing flow to obtain a latent space representation of new galaxies and feed these low-dimensional representations to the decoder of the VAE to simulate new images.
\citet{10.1093/mnras/stab1214} showed that such an architecture can be used to simulate complex and realistic models of galaxies with a hierarchical Bayesian framework to embed physical information within deep learning models. 
Although the authors explicitly account for both the PSF and the noise, in our work, we do not decouple the PSF. 
Instead, we take a more conservative approach and train and evaluate our model with a fixed PSF to increase the effect of blending and make the task more challenging for the deblender.
Our goal is to obtain a MAP estimate similar to that of \cite{https://doi.org/10.48550/arxiv.1912.03980} but instead of modeling the prior at pixel level with an auto-regressive network, we perform dimensionality reduction with a VAE and model our data-driven prior in the latent space using a normalizing flow network that provides with an explicit likelihood. 
In other words, we obtain the MAP solution in the low-dimensional latent space instead of the high-dimensional pixel space. In this section, we explain the data preprocessing,  architectures, and training of VAE and normalizing flow, and discuss the methodology of our deblender.

\subsection{Data normalization}
\label{subsec: Data normalization}

Simulated galaxies typically have a wide range of flux values in each pixel. 
Training the neural network directly with this data can make the training very unstable, so we need a mechanism to normalize the data.
Although \citet{10.1093/mnras/staa3062} have found that a non-linear normalization is ideal for deblending with VAEs, the loss function in such a normalized space is difficult to interpret.
Since we want to obtain the final MAP solution in the flux space, we choose a linear normalization for the entire dataset such that the range of flux values falls mostly between 0 and 1. 
To be more precise, we divide the flux in all pixels by a factor of $10^4$, which is sufficient to bring the range of pixel values to a level that can be handled by the network.
Studies with other normalization techniques are left as future work.

\subsection{Variational autoencoders}
\label{subsec: VAE}

An autoencoder falls under a class of neural networks that learn an encoding of high-dimensional data in a low-dimensional latent space. The architecture can be divided into three parts: the encoder, the latent space, and the decoder. 
The traditional auto-encoder provides a deterministic value while projecting data into the latent space. The loss function focuses only on the reconstruction and does not include any constraints on the properties of the latent space. 
Since the latent space is unconstrained, it is difficult to use the decoder as a generative model because we have no information about the underlying latent space distribution to sample from.  
First introduced by \cite{https://doi.org/10.48550/arxiv.1312.6114}, VAE overcomes this limitation by predicting a distribution in the latent space instead of a deterministic output and imposing a regularization over this predicted distribution. 

More formally, if $x$ represents the observed data and $z$ denotes the hidden latent variables, the decoder is represented by the mapping $p_\theta(x|z)$, while the encoder is represented by a second parametric function $q_\phi (z|x)$, where $\theta$ and $\phi$ refer to the weights of the network. 

The optimization step of the VAE involves finding the parameters $\hat \theta$ and $\hat \phi$ that minimize the evidence lower bound (ELBO) on the right-hand side of the inequality 
\begin{equation}
    \log p(x) \geq \E_{z\sim q_\phi(z|x)}[\log p_\theta(x|z)] - \D_{\rm{KL}}[q_\phi(z|x) || p(z)],
    \label{equation: ELBO}
\end{equation}
where $\log p(x)$ is called the evidence, the first term is the expectation of the likelihood, $p(z)$ is the latent space prior, and $\D_{\rm{KL}}$ represents the Kullback–Leibler (KL) divergence which measures how different the two distributions are.

Optimization is done by minimizing the ELBO loss function. 
However, as observed very frequently in VAE literature \citep{chen2017variational}, the KL divergence term in ELBO dominates and prevents the network from properly reconstructing galaxies. 
Among several different approaches, typically annealing-based methods are used to slowly increase the weight of the KL divergence term during training. 
Alternatively, this dominance can also be mitigated by multiplying the KL term with another hyperparameter $\beta$ in the loss function, 
to adjust the relative weight between the reconstruction and the regularization terms.
These models are called the $\beta$-VAE \citep{DBLP:conf/iclr/HigginsMPBGBML17}.

After lowering the weight of KL divergence, we observed that the reconstructions were still poor in the \textit{u}-band. 
This is because the  VAE starts learning features in the \textit{u}-band only in later iterations due to the relatively lower S/N of galaxies compared to the other bands.
To address this challenge, we include a warm-up phase in the VAE training with a slightly modified loss function.
We introduce the structural similarity index measure \cite[SSIM,][]{1284395} in a way that the neural network is forced to predict correct structural information in all the bands from the initial iterations.
The SSIM is a metric typically used in image processing to assess the similarity between two images. 
An SSIM value close to $1$ represents highly correlated structural information between the two images, while $-1$ represents anti-correlation.
Using SSIM, we define the warm-up loss as

\begin{equation}
    \begin{split}
    \Lagr = - \E_{z\sim q_\phi(z|x)}[\log p_\theta(x|z)(1 - \alpha \ssim)] + \beta \D_{\rm{KL}}[q_\phi(z|x) || p(z)],
    \end{split}
    \label{equation: ELBO warm-up}
\end{equation}
where $\ssim$ denotes the average SSIM across filters, computed using input and output images of each band normalized to a maximum value of 1, and $\alpha$ is a weight applied to SSIM. In practice, we set $\alpha$ in a way that it reduces to zero after a few initial iterations, as described in Appendix \ref{subsec: VAE HyperParameters}. The exact architecture of the VAE and hyperparameters of the model used to obtain our results is presented in Appendix \ref{subsec: VAE HyperParameters}.

Once trained, the VAE can be used as a generative model by sampling from the latent space prior $p(z)$ in the ELBO, and feeding it to the decoder.
However, this assumes that the prior is representative of the aggregate posterior which is not strictly valid for a VAE, and introducing a $\beta<1$ in the $\beta-$VAE makes the aggregate posterior depart further from the prior.
Therefore, sampling from the prior would lead to biased and out-of-distribution samples.
\citet{2017arXiv171105772E} proposed using GANs to model the latent space to generate high-fidelity images. GANs, however, do not provide a way to evaluate the likelihood necessary for the MAP estimate.
As an alternative to GANs, in Section \ref{subsection: Normalizing Flows} we discuss in detail another generative model called normalizing flows that gives access to a tractable likelihood.

\subsection{Normalizing flows}
\label{subsection: Normalizing Flows}

Normalizing flows are capable of modeling a complex distribution by sequentially applying bijective transformations to a simpler distribution that can easily be sampled.
Mathematically, if we consider $\X$ and $\Z$ to be two random variables related by bijective mapping $f_\varphi : \mathbb{R}^n \rightarrow \mathbb{R}^n$, parameterized by $\varphi$, such that $\X$ = $f_\varphi(\Z)$.
The goal of fitting the normalizing flow is to find appropriate parameters $\varphi$ so that the transformed distribution is similar to the target distribution we are trying to model, which in our case is the aggregate posterior in the VAE latent space. 

In our work, we used a family of normalizing flows, called masked autoregressive flows (MAFs), presented by \citet{https://doi.org/10.48550/arxiv.1705.07057}, which are typically parameterized by a dense network. 
However, one layer of MAF is not expressive enough to model a non-trivial distribution, and results depend heavily on the order of variables.
Thus, we must apply a set of such transformations to the base distribution, while performing permutations to the order of variables between layers, i.e.
\begin{equation}
    \x = f_\varphi(\z) = f_{\varphi_n} ... \circ f_{\varphi_2} \circ f_{\varphi_1}(\z)\,,
\end{equation}
where $f_{\varphi_j}$ represents the $j^{\text{th}}$ layer of transformation, parameterized by a neural network $\varphi_j$.

We train the model by minimizing the negative log-likelihood of the latent space representations of galaxies in \IsolatedSims obtained by a forward pass through the trained encoder as shown in Figure \ref{fig: normalizing flow}. Appendix \ref{subsec: normalizing flow HyperParameters} describes the network architecture and training hyperparameters for the model used to generate the results hereafter.

\begin{figure}
\centering
\includegraphics[width=.5 \textwidth]{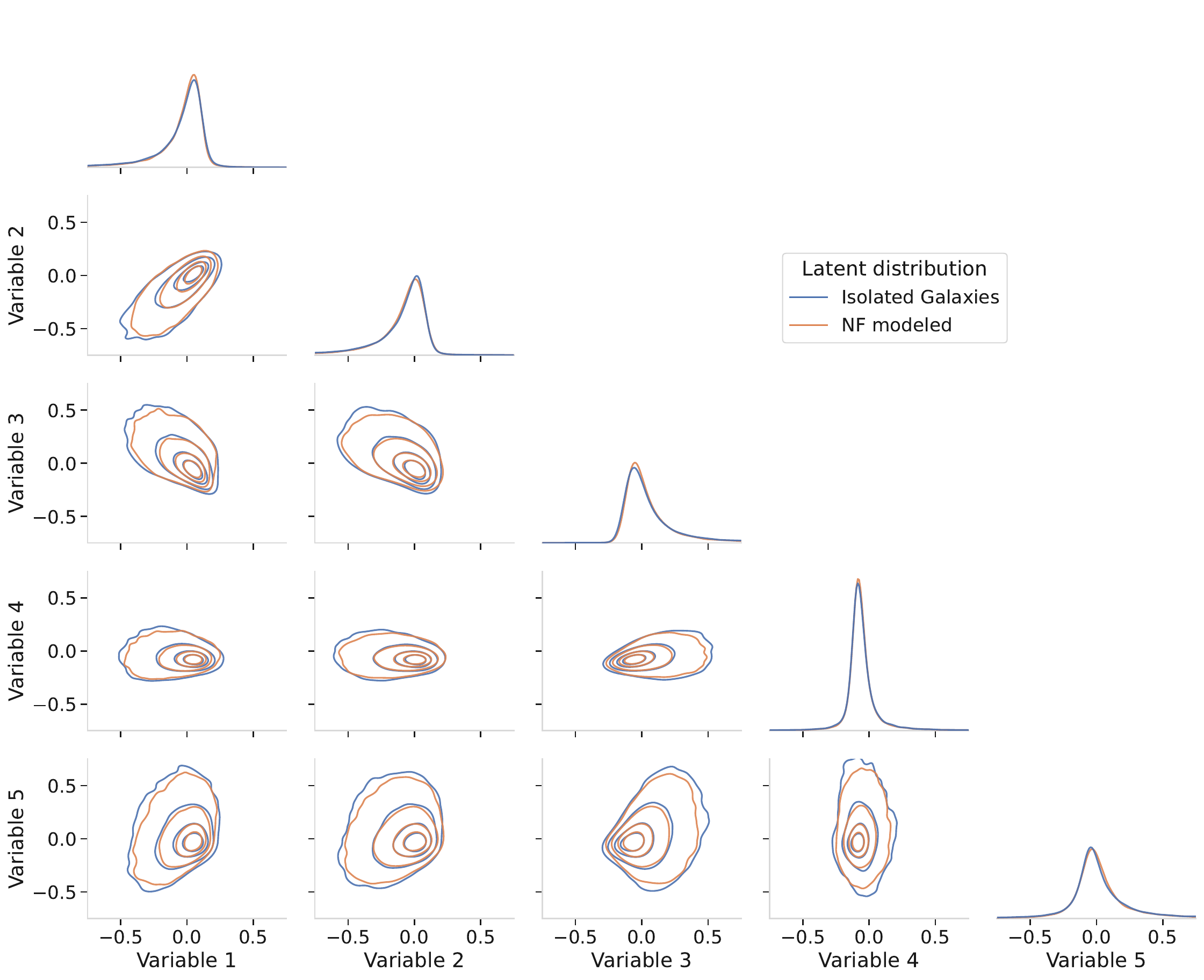}
\caption{Comparison of the latent space distribution of \num{50000} samples each from the validation set of \IsolatedSims\ and the distribution learned by the normalizing flow. As one moves out from the center of the distribution, the contours represent $25\%$, $50\%$, $75\%$, and $90\%$ of the probability mass.}
\label{fig: normalizing flow distribution correlation}
\end{figure}

\begin{figure}
     \centering
     \subfloat[]{
         \centering
         \includegraphics[width=.49\textwidth]{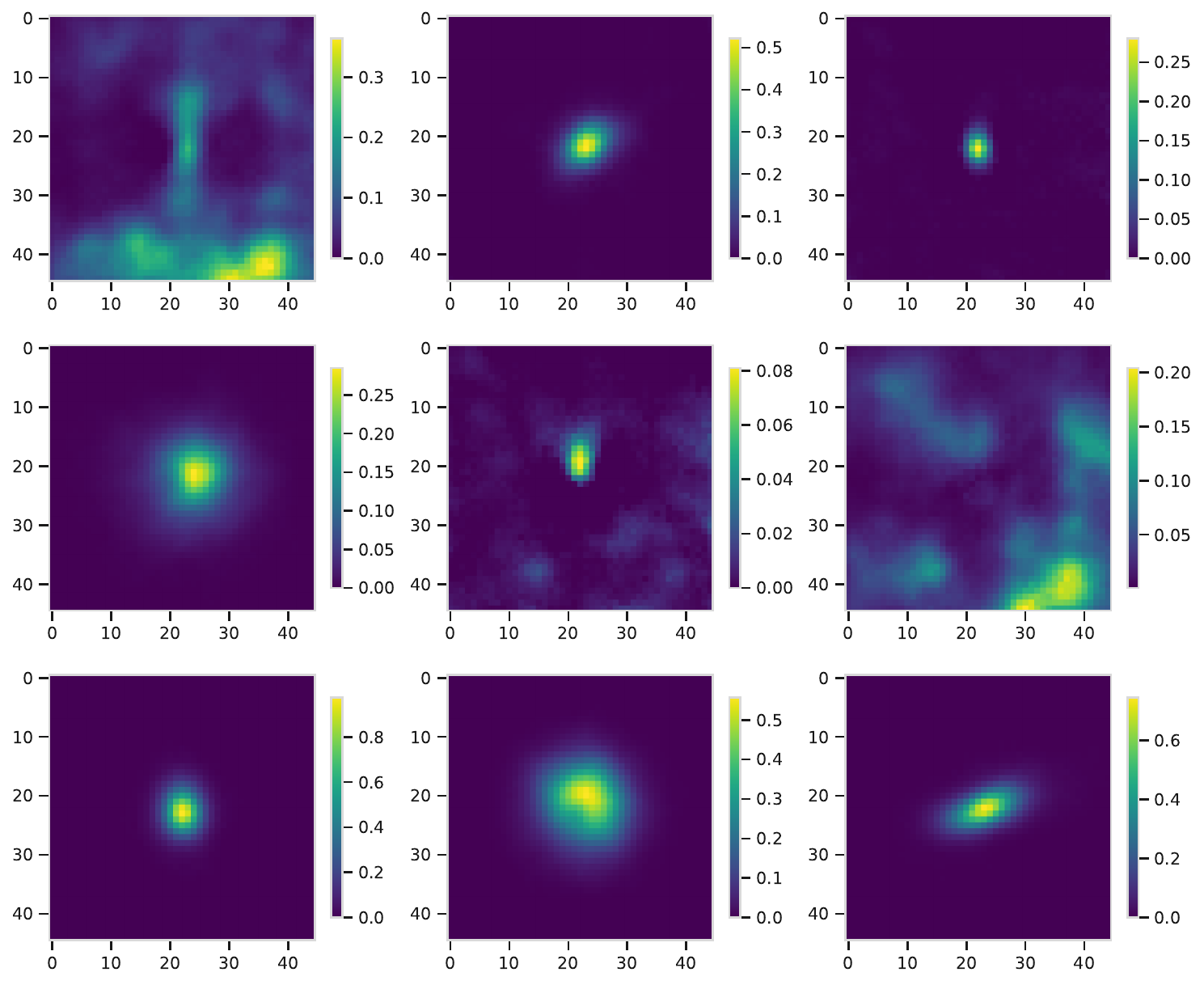}
         \label{fig: VAE simulations}
     }
     \hfill
     \subfloat[]{
         \centering
         \includegraphics[width=.49\textwidth]{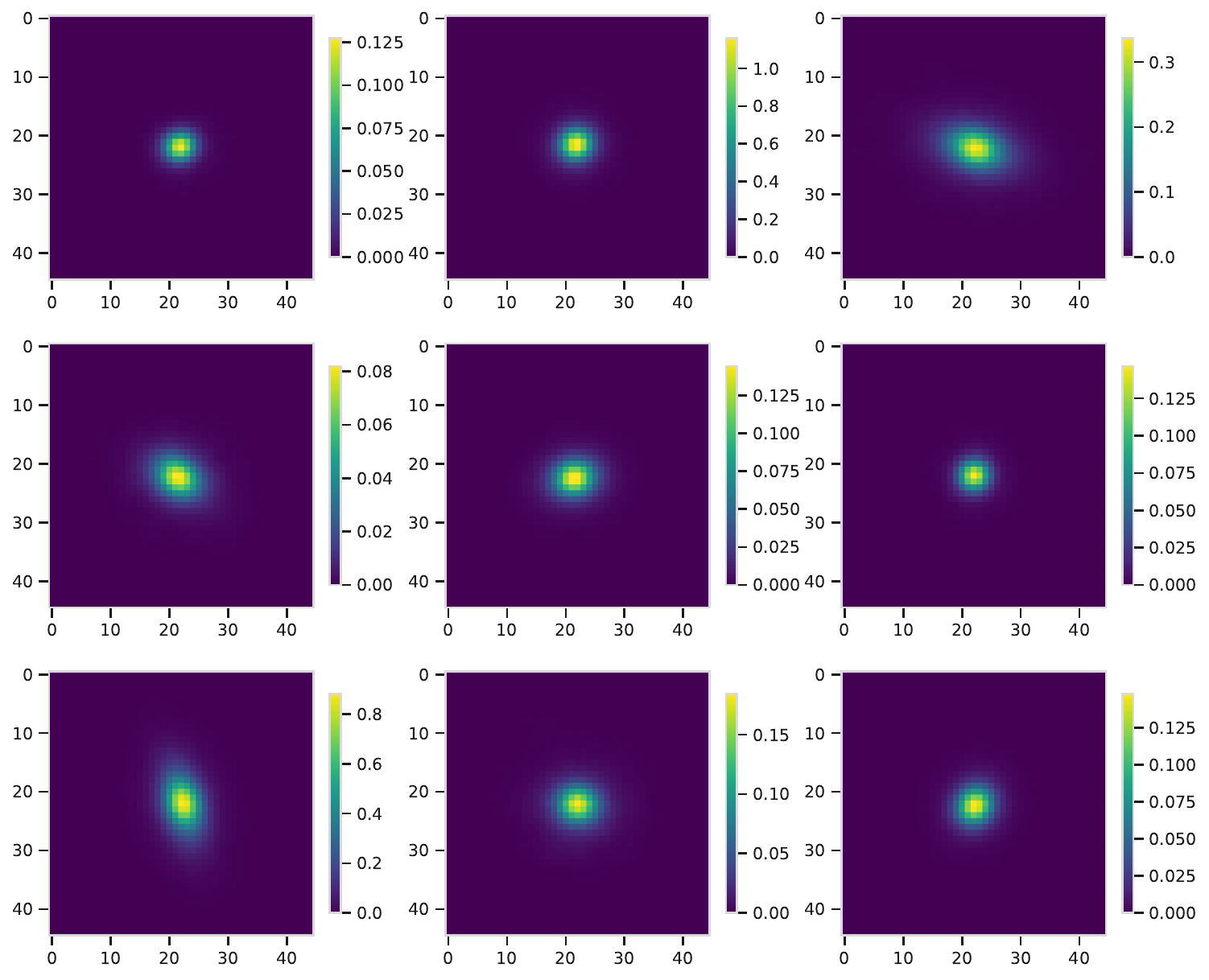}
         \label{fig: normalizing flow simulations}
    }

    \caption{Examples of randomly chosen single galaxy simulations in the \textit{r}-band using the trained decoder as the generative model. Panel (a) shows galaxies simulated with VAE after sampling from the ELBO prior, while panel (b) shows galaxies simulated after sampling from the distribution learned using the normalizing flow. The color bar for each plot shows the normalized electron counts in the pixels.}
    \label{fig: generative model simulations}
\end{figure}

In Figure \ref{fig: normalizing flow distribution correlation}, we show that the trained normalizing flow successfully models the underlying latent space representation of the galaxies. 
Figure \ref{fig: generative model simulations} shows the impact on the simulation of galaxies with the VAE when shifting to a normalizing flow instead of the ELBO prior.
We can conclude from Figure \ref{fig: generative model simulations} that sampling from the standard normal prior used to train the VAE fails to generate realistic-looking galaxies, but sampling from the aggregate posterior as modeled by the normalizing flow gives more reliable simulations 
\citep{10.1093/mnras/stab1214, arcelin:tel-03553937}.
We now use this architecture to obtain the log prior required for the MAP solution.

%-------------------------------------------------------------------
\subsection{VAE-deblender}
\label{subsection: deblender}

\begin{figure}
\centering
\includegraphics[width=.5\textwidth]{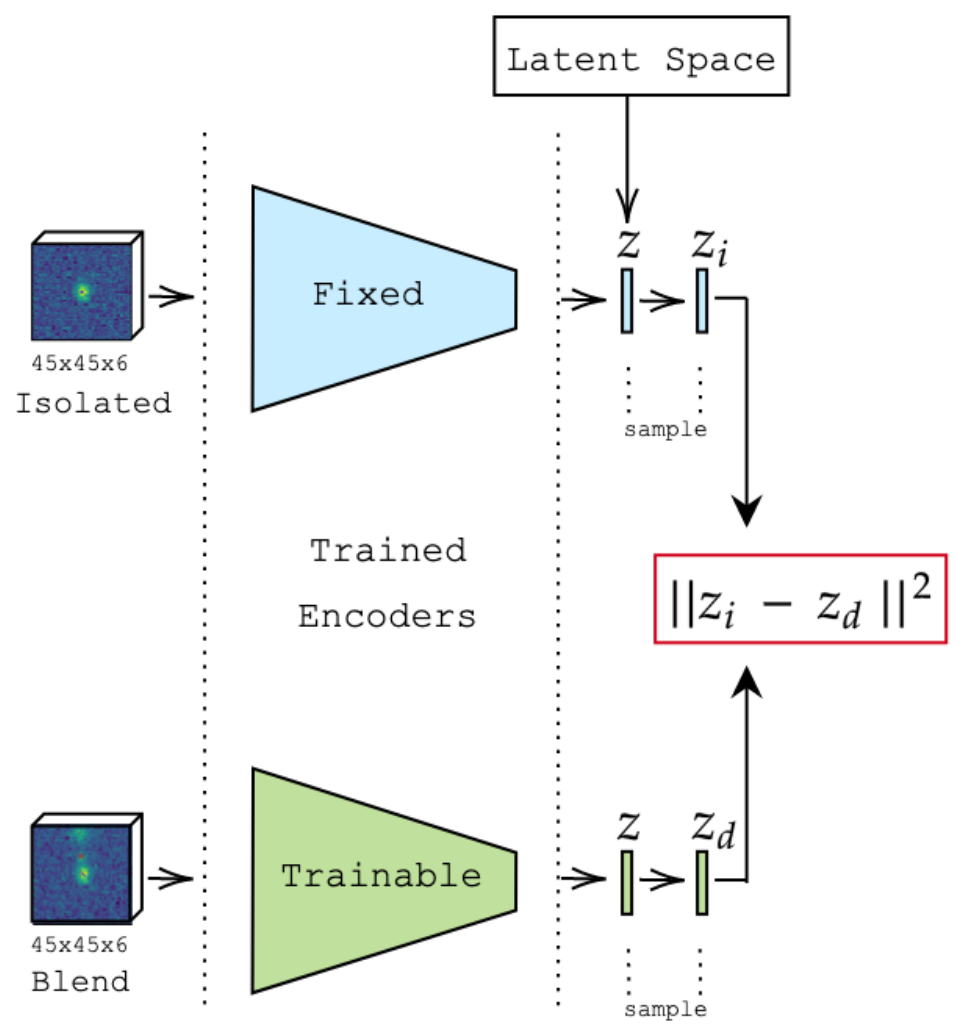}
\caption{Schematic of the training of the \VAEdeblender using two copies of the encoder network illustrated in Figure \ref{fig: VAE diag}. The trainable network learns to predict the latent space representation of the central galaxy by minimizing the Euclidian distance (marked in red) between its prediction ($z_d$) and the representation of the isolated galaxy ($z_i$) obtained by the fixed encoder trained as described in Appendix \ref{subsec: VAE HyperParameters}.}
\label{fig: Modified VAE training}
\end{figure}

\begin{figure}
\centering
\includegraphics[width=.5\textwidth]{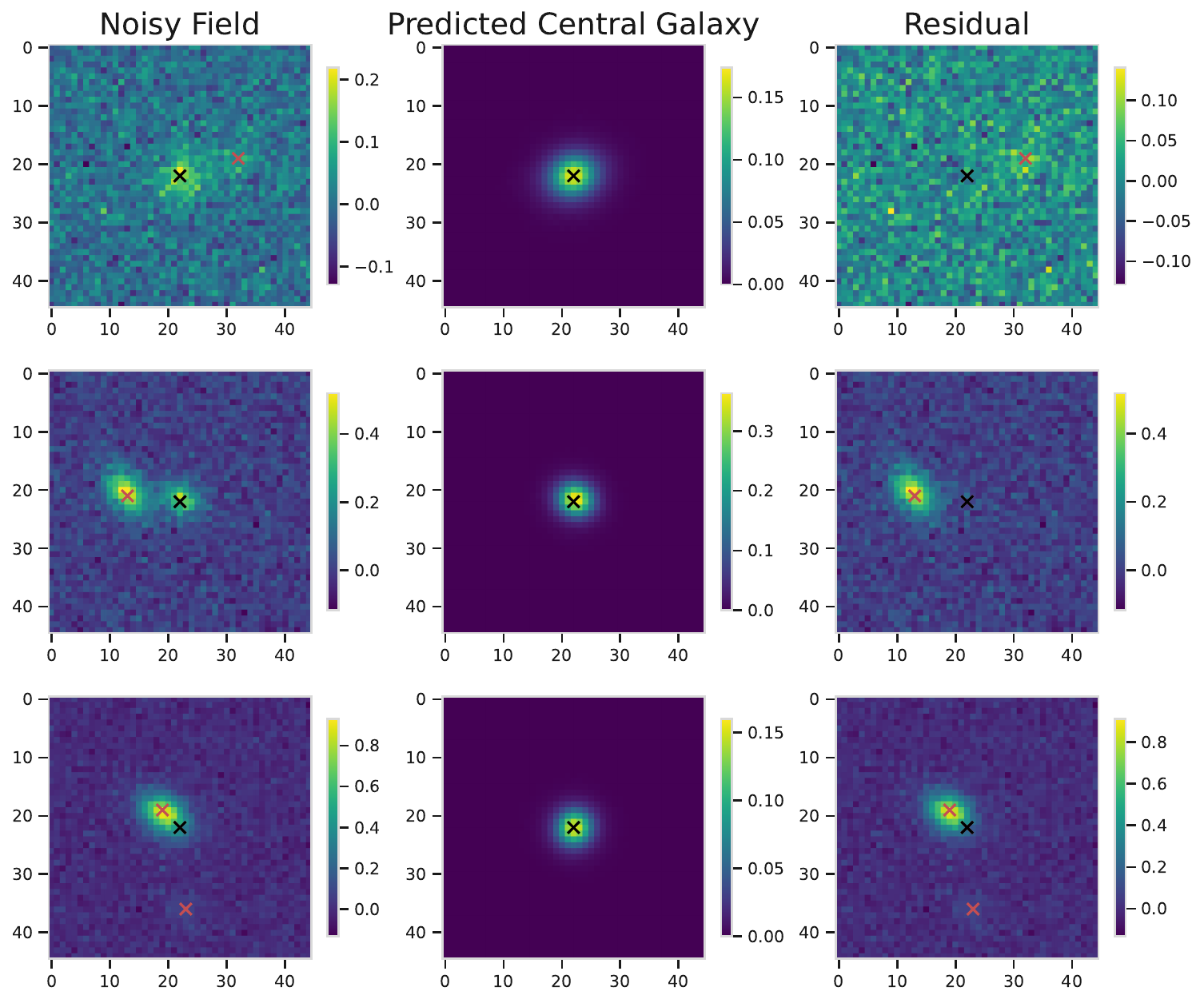}
\caption{Examples of deblended central galaxies from the validation set of \BlendedSims, predicted in the \textit{r}-band by a single forward pass through the \VAEdeblender. Each row represents the same scene and the first column shows the blended scene; the second column shows the predicted model for the central galaxy; and the third column is the result of subtracting the predicted model from the blended field. The color bar for each plot shows the normalized electron counts in the pixels.}
\label{fig: Modified VAE deblending}
\end{figure}

Once we have the normalizing flow and trained VAE, it is possible to generate new galaxies by sampling from the distribution learned from the normalizing flow and feeding it to the decoder. 
The goal hereafter is to obtain the MAP solution in the latent space with gradient descent. 
To reduce the time taken for optimization, it is essential to provide the model with a reasonable initialization of latent points. 

For this, we adopt the idea presented by \citet{10.1093/mnras/staa3062}, where the authors retrain only the encoder as a deblender, to find the latent space representation of the central galaxy in postage stamps containing blends. 
While keeping the pre-trained decoder fixed, the authors provide blended stamps as input to the encoder and compute the reconstruction loss of the decoder output with the isolated central galaxy in the input stamp.
Upon minimizing the negative log-likelihood under this setting, the newly retrained encoder outputs the amortized posterior distribution of only the central galaxy in the latent space.
In other words, the encoder part of the VAE acts as a deblender, extracting only the central galaxy in the postage stamp. 
Since the decoder is fixed, this step does not affect the underlying distribution of isolated galaxies and does not require retraining the normalizing flow.

Although we follow a similar ideology as \citet{10.1093/mnras/staa3062}, the loss function is modified to accommodate our choice of a simpler network architecture and linear data normalization. 
In our case, retraining the encoder as a deblender is more difficult because of the choice of linear normalization. 
So, instead of computing the loss in the image space, we shift to the latent space. 
For this purpose, we start with two sets of the trained encoder as shown in Figure \ref{fig: Modified VAE training}. 
Using the \BlendedSims we feed images of an isolated galaxy to the first encoder, and to the second encoder we provide a corresponding blended scene and the same central galaxy. 
Let $z_i$ (isolated representation) and $z_d$ (deblended representation) be samples from the amortized posterior obtained from the first and second encoders, respectively.
Keeping the weights of the first encoder fixed, we retrain only the second encoder to minimize the L2-norm or Euclidean distance between $z_i$ and $z_d$ given by
\begin{equation}
    \texttt{L2-norm} = ||z_i - z_d||^2,
\end{equation}
and the choice of training hyperparameters used is presented in Appendix \ref{subsec: VAE-Deblender HyperParameters}.

Once trained, the second network can be combined with the previously trained decoder and used as a deblender similar to \cite{10.1093/mnras/staa3062} as shown in Figure \ref{fig: Modified VAE deblending}. 
Although this architecture predicts a distribution in the latent space, we concern ourselves with only the mean of this distribution; hereafter,  we call this the \VAEdeblender.
In our algorithm, we used the \VAEdeblender only to initialize the latent space before the gradient descent for the MAP estimate.

Given a field of blended galaxies, we can extract postage stamps centered around each detection, feed them to the second encoder, and obtain an initial prediction for the latent space representation of the central galaxy by sampling from the amortized posterior approximated by the retrained encoder.

\subsection{Deblending}
\begin{figure*}
     \centering
    \includegraphics[width=.9\textwidth]{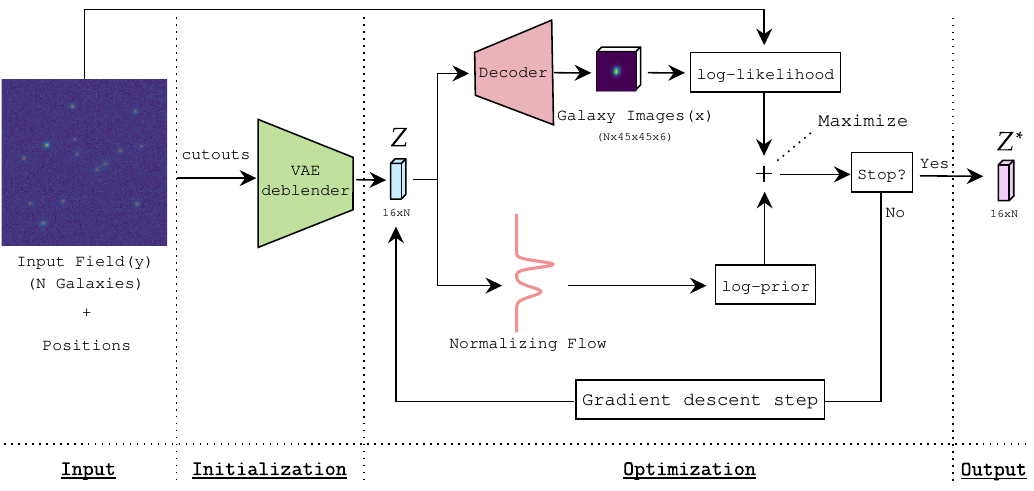}
    \caption{Schematic of \MADNESS workflow. The algorithm takes as input a field of galaxies and their detected positions. 
    The \VAEdeblender consisting of convolutional and dense layers initializes the latent space ($Z$), which is followed by the optimization of the sum of the log-prior and the log-likelihood by using the decoder (dense and transposed convolutional layers) to map latent space representations to the image space and the normalizing flow which uses MAF layers to compute the log-prior.
    Once the stopping condition is satisfied based on the rate of decrease of the minimization objective, we obtain the maximum a posteriori solution $Z^*$ as output. 
    The corresponding image space solutions can be found with a forward pass through the decoder.}
    \label{fig: MADNESS Schematic}
\end{figure*}

The VAE has already proved to be a powerful tool in the domain of generative models.
However, it fails to provide an explicit likelihood.
By combining the VAE with the flow network, the so-called Flow-VAE can be used to obtain the MAP solution of the deblending problem. 
Say we have a field of galaxies $y$, and our goal is to obtain a set of galaxies $X$ such that 
\begin{equation}
    \hat{X}= \argmax_X  [p(y|X) p(X)],
    \label{MAP}
\end{equation}
where $p(y|X)$ is the data likelihood and $p(X)$ is the prior representing the probability of the reconstructions to look like galaxies.
As the prior is hard to evaluate in the pixel space without using symmetry constraints as in \scarlet, we turn towards the latent space.
Since the normalizing flow has been trained to model the aggregate posterior of the VAE, we can evaluate the prior term in the latent space using the distribution modeled by the normalizing flow.
Therefore, Equation \ref{MAP} is rewritten as
\begin{equation}
    \hat{Z}= \argmax_Z  [p_\theta(y|Z) p_\varphi(Z)],
    \label{equation: shift to latent space}
\end{equation}
where $Z$ is the set of latent space representations of galaxies, and $\theta$ and $\varphi$ are the pre-trained weights of the VAE and normalizing flow, respectively. We can rewrite Equation \ref{equation: shift to latent space} as the minimization of
\begin{equation}
    \hat{Z}= \argmin_Z \left [- \log p_\theta(y|Z) - \log_\varphi p(Z) \right ].
    \label{equation: deblending solution}
\end{equation}

The caveat here is that since we have shifted to the latent space for the minimization, we cannot evaluate the log-likelihood directly without mapping to the image space. 
As the VAE was trained to generate images, the decoder is essentially the mapping back to the image space. 
To evaluate the log-likelihood term, we used the trained decoder $p_\theta$ to go from the latent space representation of each galaxy to the image space as shown in Figure \ref{fig: MADNESS Schematic}. 
Given the architectural choice of $45 \times 45$ pixel cutouts used to train the VAE, the final predicted image of each galaxy is a cutout of the same dimension.
By summing over each cutout at the corresponding detection location of the galaxy, we obtain the reconstructed field $\fieldX$ given by
\begin{equation}
    \fieldX =  \sum_{\forall z_i \in Z} \rm{PAD}(p_\theta(z_i), j_i, k_i),
    \label{eqn: reconstructed field}
\end{equation}
where $z_i$ is the representation of the $i^{\rm{th}}$ galaxy in the field, $p_\theta$ is the pre-trained decoder network of the VAE, and the padding function $\rm{PAD}$ changes the dimensions of the prediction to the size of the field based on the pixel coordinates ($j_i$, $k_i$) of the center of the galaxy in the field.

Assuming an uncorrelated Gaussian approximation to Poisson noise in each pixel, we can now compute the log-likelihood term using the reconstructed field as
\begin{equation}
    \log p_\theta(y|Z) =  - (y - \fieldX)^T \Sigma^{-1} (y - \fieldX),
    \label{eqn: log-likelihood}
\end{equation}
where $p_\theta$ is the decoder and $\Sigma$ is the covariance matrix, which reduces to a diagonal matrix as the noise is assumed to be uncorrelated. Putting everything together, we obtain
\begin{equation}
    \hat{Z}= \argmin_Z \xspace \left[ (y - \fieldX)^T \Sigma^{-1} (y - \fieldX) - \sum_{\forall z_i \in Z} \log p_\varphi(z_i) \right].
    \label{equation: final deblending solution}
\end{equation}

We used the Adam optimizer for the gradient descent with an initial learning rate of $0.05$ and a scheduler that reduces it by a factor of $0.8$ every $30$ epochs. 
The optimization is terminated when the decrease in the exponentially weighted moving average of the final $15$ steps drops below a factor of $0.05$ of that of the initial $15$ steps.

Figure \ref{fig: MADNESS Schematic} summarizes the workflow of the algorithm. After the initialization of the latent space with the \VAEdeblender,
the decoder allows one to evaluate the first term of Equation \ref{equation: final deblending solution} while the normalizing flow allows the evaluation of the log-prior for each galaxy.
For each field, we obtain the MAP solution $\hat{Z}$ in the latent space for all the galaxies simultaneously by minimizing the objective in Equation \ref{equation: final deblending solution} for all galaxies until the stopping condition is satisfied.
Once the solution $\hat{Z}$ is obtained, the corresponding image space solution is obtained by a forward pass through the decoder.

\section{Results}
\label{sec:Results}

Once we obtain the MAP estimate in the VAE latent space, we use the \TestFields described in \ref{subsection: Test Data} to test the reconstructions in terms of the total flux, \textit{g-r} color, shapes, and morphology indicators. As a sanity check, we verify that the MAP optimization improves the results from the \VAEdeblender and we further compare \MADNESS against \scarlet, using the configuration described in Appendix \ref{section: appendix scarlet config}.

These tests were run on single V100 GPUs at CC-IN2P3, where we parallelized the computation of 20 fields from the \TestFields\ dataset at a time and observed a gradient descent convergence within 1 second per field (approximately 16 galaxies/sec). 

Although in a more realistic scenario, the deblenders will be used alongside a detection algorithm, here we remove detection biases and report the scores for all the metrics, providing the true centers of all galaxies to the deblenders.

\subsection{Metrics}
\label{subsec: Metrics}

We evaluated the performance of the flux reconstruction using aperture-photometry that computes flux within a given radius around the true center of the galaxies. We compute the \textit{g-r} color of the galaxies, which is the difference in magnitude in the \textit{g} and \textit{r} bands, to ensure that the cross-calibration between bands is preserved. 

To test the morphology of the reconstructions, we computed the SSIM described in Section \ref{subsec: VAE} and the cosine similarity, which is the cosine of the angle between the reconstructed model and the ground truth flattened as a 1D vector. 
We characterize the shapes of the galaxies with ellipticities obtained from adaptive moments \citep[see][for a detailed discussion on adaptive moments]{2003MNRAS.343..459H}, computed using the \galsim \texttt{HSM}  module.

The results for these metrics are expected to vary significantly based on the S/N and blending of galaxies.
Therefore, to properly understand their impact, we used the following metrics:

\begin{enumerate}

    \item the \textit{blendedness} \citep{10.1093/pasj/psx080} metric quantifies the effect of blending for each galaxy.
    For a galaxy in the field, the blendedness in the $n^{\rm{th}}$ filter is defined as:
    \begin{equation}
        \beta_n = 1 - \frac{S_n \cdot S_n}{S^{b}_n \cdot S_n},
    \end{equation}
    where $S_n$ is the vector of true pixel values for the galaxy and $S^b_n$ is the flux vector from the blended field. 
    For heavily blended objects in the limit of $S^b_n \gg S_n$ where $S_n$ is non-zero, $\beta_n$ tends to 1.
    For the other extreme case where the effect of blending is negligible, $S^b_n \to S_n$ near the galaxy so $\beta_n$ approaches 0. 
    For a more visual interpretation, Appendix \ref{section: appendix metrics} shows cutouts of galaxies, as the S/N and blendedness vary.

    \item \textit{contamination} ($\kappa$) of a galaxy is defined as the relative residual in aperture-photometry when measured in a blended scene, compared to the isolated case. In a given galaxy, the contamination for $n ^{\rm{th}}$ band is given by
    \begin{equation}
        \kappa_{n} = \frac{F_n^b}{F_n} - 1,
    \end{equation}
    where $F_n^b$ and $F_n$ are the aperture-photometry measured flux for the blended and isolated galaxy, respectively.
    Appendix \ref{section: appendix metrics} gives a visual representation of how galaxies look as we vary $\kappa$ in different S/N ranges.

    \item S/N:
    In Section \ref{sec:Dataset}, a selection cut of $\text{S/N} \geq10$ was applied on the \textit{r}-band S/N of each galaxy, computed at the catalog level.
    For consistency, we continue with the same definition of S/N to evaluate our results.

\end{enumerate}

\subsection{Errors in aperture-photometry of deblended models}
\label{subsec: Aperture Photometry}

\begin{figure}
     \centering
     \subfloat[]{
         \centering
         \includegraphics[width=.49\textwidth]{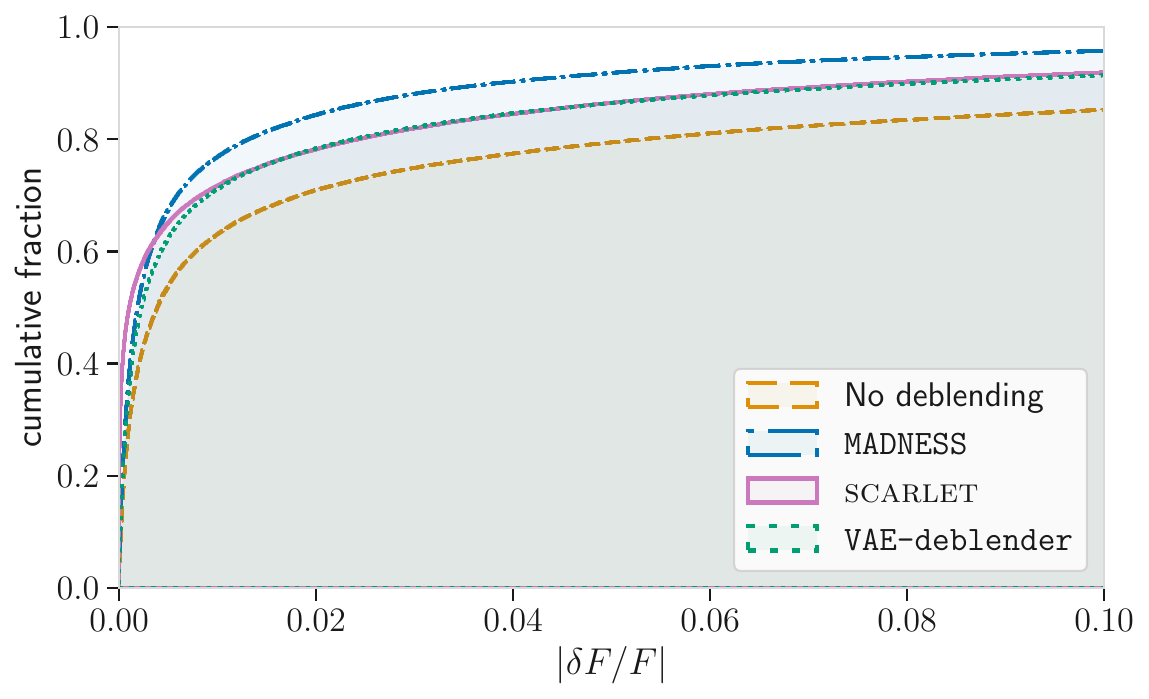}
         \label{fig: cumulative distribution of the aperture-photometry errors}
     }
     \hfill
     \subfloat[]{
         \centering
         \includegraphics[width=.49\textwidth]{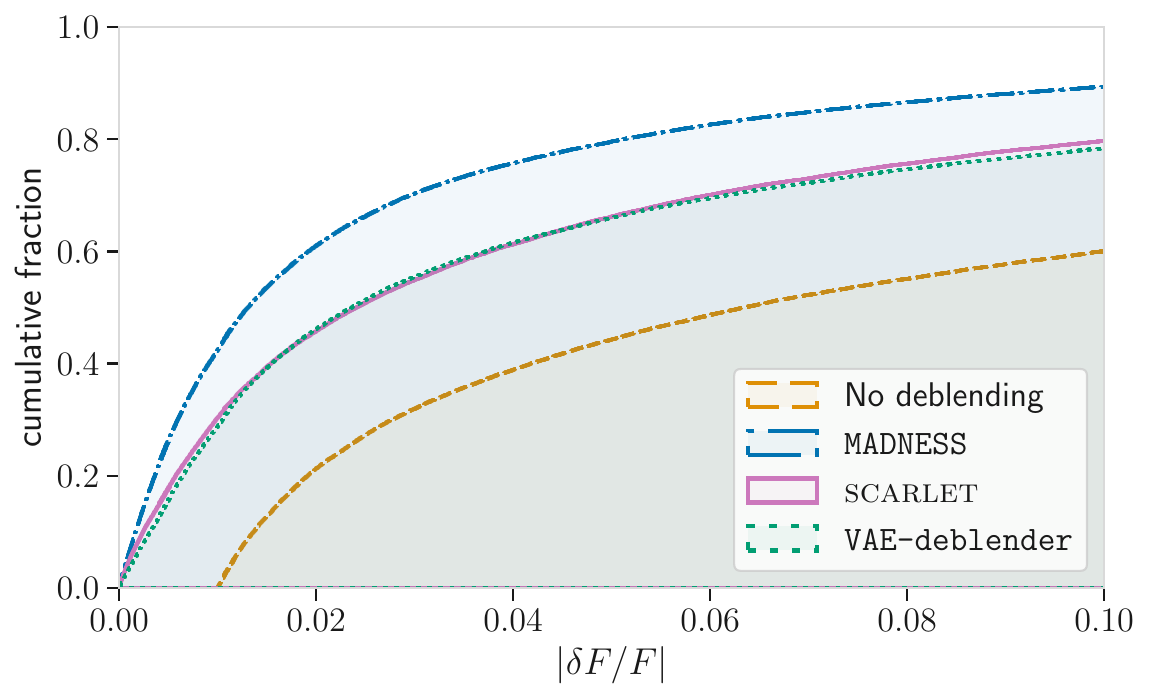}
         \label{fig: cumulative distribution of the aperture-photometry errors contamination cut}
    }

    \caption{Comparison of the absolute value of relative residuals $| \delta F/ F |$ in aperture-photometry among \MADNESS, \scarlet, \VAEdeblender, and the original field without deblending.
    Panel (a) shows the cumulative distribution of the absolute value of the relative residual in the \textit{r}-band for the entire \TestFields dataset, 
    while panel (b) shows the distribution of $| \delta F/ F |$ for blended galaxies with $\kappa_r>0.01$, which represents at least $1\%$ flux contribution from neighboring sources.}
    \label{fig: aper phot distrib}
\end{figure}

\begin{figure}
     \centering

    \includegraphics[width=.49\textwidth]{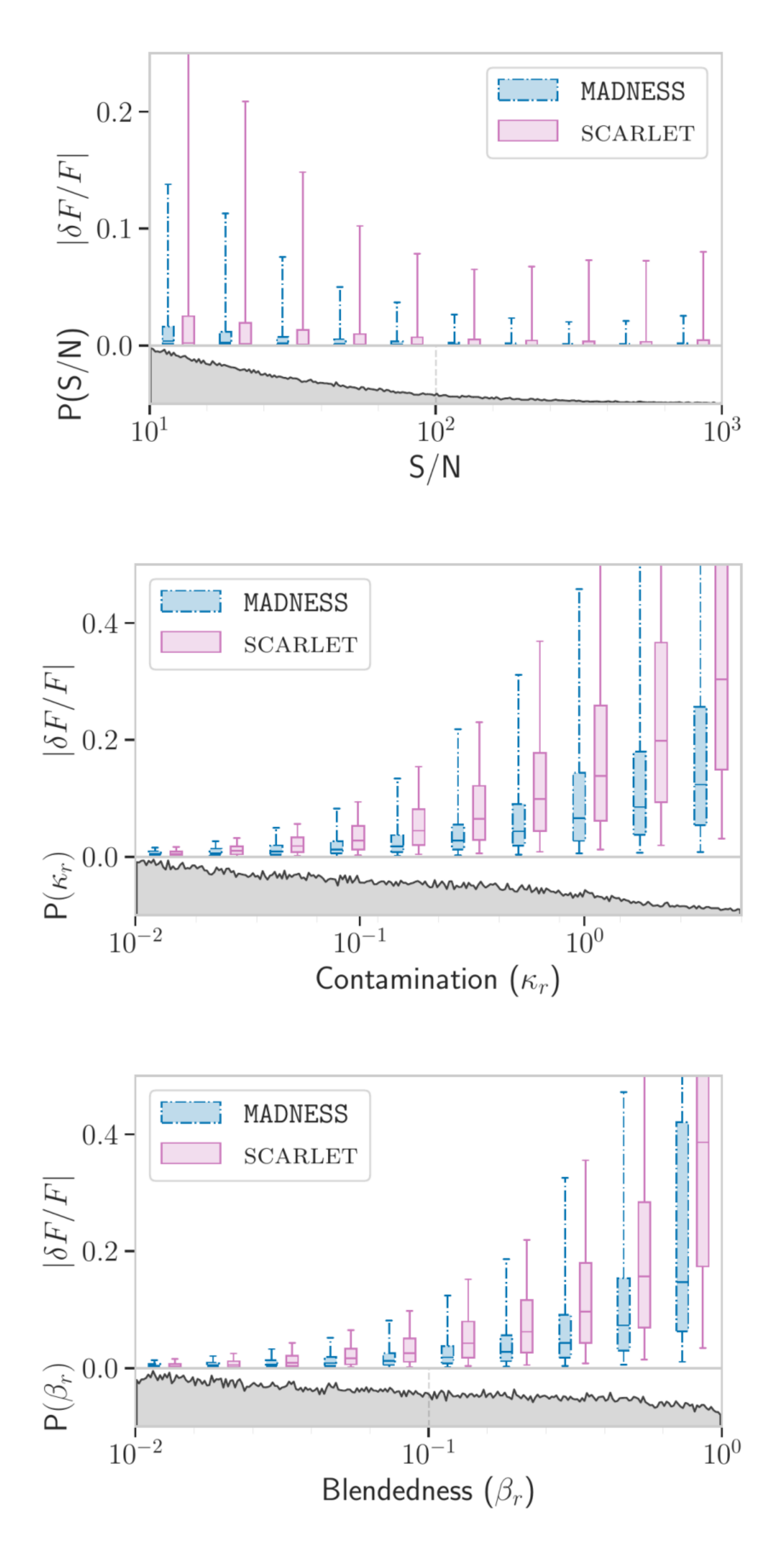}
    \caption{Aperture-photometry results in the $r$-band of galaxies from \TestFields after deblending with \MADNESS, \scarlet, in terms of signal-to-noise ratio (S/N), blendedness ($\beta$), and contamination ($\kappa$) in the \textit{r}-band. 
    In each panel, the top plot shows the distribution of the absolute value of relative flux residual ($|\delta F/F|$) while the bottom plot shows the distribution of galaxies for the metric on the x-axis.
    The boxes in the plot on top show the range from the $25 ^{\rm{th}}$ to $75 ^{\rm{th}}$ percentile, with a horizontal line representing the median, while the whiskers represent the $5 ^{\rm{th}}$ to $95 ^{\rm{th}}$ percentile of relative residuals in the \textit{r}-band.}
    \label{fig: box plot aperture phot}
\end{figure} 

For both \MADNESS and \scarlet, we compute the aperture-photometry for each galaxy by centering on the ground truth position and subtracting the pixel values of all other predicted galaxies from the field except the central one.
We choose elliptical apertures with semi-major and semi-minor axes $a$ and $b$ for each band given by
\begin{equation} 
    a = \sqrt{(2a_{hlr})^2 + d^2}~, ~ \rm{and}~~ b = \sqrt{(2b_{hlr})^2 + d^2},
\end{equation}
where $a_{hlr}$ and $b_{hlr}$ are the half-light radii of the galaxy and $d$ is the PSF full-width-half-maximum (FWHM) of the given band.

 Using the \SourceExtractor library \citep{Barbary2016}, we computed the flux of the reconstructions in each band. Figure \ref{fig: cumulative distribution of the aperture-photometry errors} shows the absolute value of the relative flux residual, which is the ratio of the flux residual ($\delta F$) to the actual flux of the isolated galaxy. Mathematically, the relative flux residual is defined as
\begin{equation}
        \frac{\delta F}{F} = \frac{F^d-F}{F},
\end{equation}
where $F^d$ and $F$ are the $r$-band aperture-photometry flux in the same noise field for the deblended and isolated galaxy, respectively.
To provide a reference for comparison, we also computed the residuals obtained directly on the blended field, i.e., without using a deblender.

For a quantitative comparison, we computed the mean of the absolute value of the relative residuals. 
However, since the mean can be affected by outliers, we removed severely blended galaxies with contamination $\kappa_r > 5$
and observed that the mean of the absolute value of the relative residuals in the blended case is $0.142$ which implies on average, the residuals are around $14.2\%$ of the flux value.
After deblending, we significantly reduce this average to $0.020$ for \MADNESS, $0.028$ for \scarlet, and $0.039$ for the \VAEdeblender. 
So the average of \MADNESS is roughly $29\%$ and $50\%$ lower than that of \scarlet and \VAEdeblender, respectively.

The cumulative distribution of galaxies as a function of $|\delta F/F|$ in Figure \ref{fig: cumulative distribution of the aperture-photometry errors} shows that by performing gradient descent in the latent space, \MADNESS significantly improves the initial predictions of the \VAEdeblender.
Moreover, for small values of $|\delta F/F|$, \scarlet outperforms \MADNESS, but the latter outperforms the former for $|\delta F/F|\gtrapprox 0.005$. 
The better performance of \scarlet for smaller values of $|\delta F/F|$ can be attributed to the fact that \scarlet predicts models in the PSF deconvolved state, so a large fraction of galaxies that appear as blended for \MADNESS are isolated sources for \scarlet.
To test this hypothesis, we show in Figure \ref{fig: cumulative distribution of the aperture-photometry errors contamination cut}, $| \delta F/F |$ for galaxies with $\kappa_r \geq 0.01$, i.e., at least $1\%$ flux contribution from neighboring sources, that even for low values of $|\delta F/F|$, \MADNESS has a higher density of galaxies than \scarlet. 
This confirms our hypothesis that the better performance of \scarlet for small $|\delta F/F|$ is due to very slightly blended galaxies, which most likely appear isolated for \scarlet after PSF deconvolution.

In Figure \ref{fig: box plot aperture phot}, we compare the performance of the two deblenders regarding the recovered photometry as a function of S/N, blendedness, and contamination.
From these plots, we conclude that \MADNESS provides better flux reconstruction than \scarlet, especially for heavily blended galaxies; the results are consistent even in different S/N bins.

In Appendix \ref{section: appendix photometry results}, we show that our results generalize to other bands except for a deviation in the \textit{u}-band where \MADNESS and \VAEdeblender's performance drops due to the increased noise levels.

\subsection{Bias in color for deblended galaxies}
\label{subsec: color}

\begin{figure*}
     \centering
     \subfloat[]{
         \centering
         \includegraphics[width=.475\textwidth]{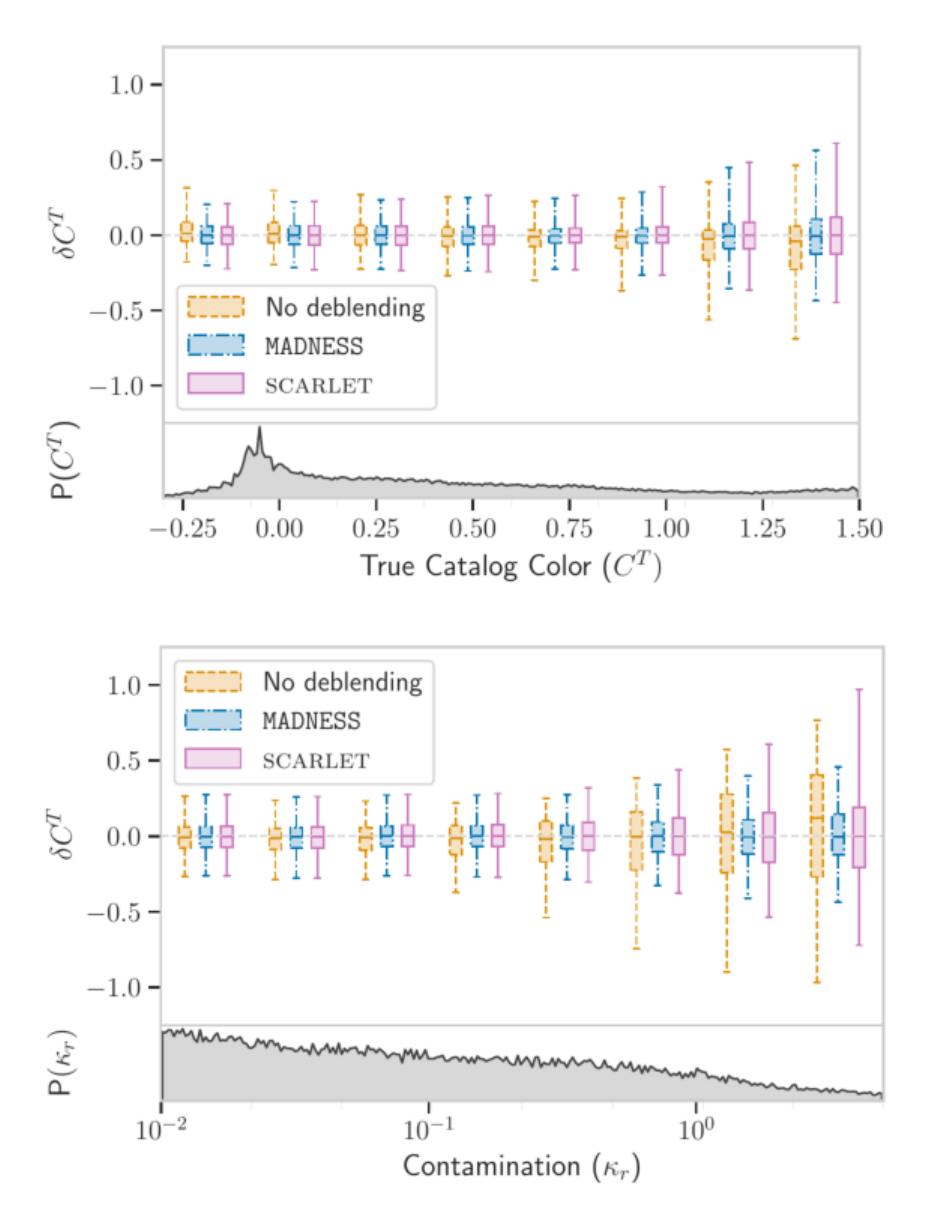}
         \label{fig: Photometry Bias}
     }
     \hspace{1em}
     \subfloat[]{
         \centering
         \includegraphics[width=.475\textwidth]{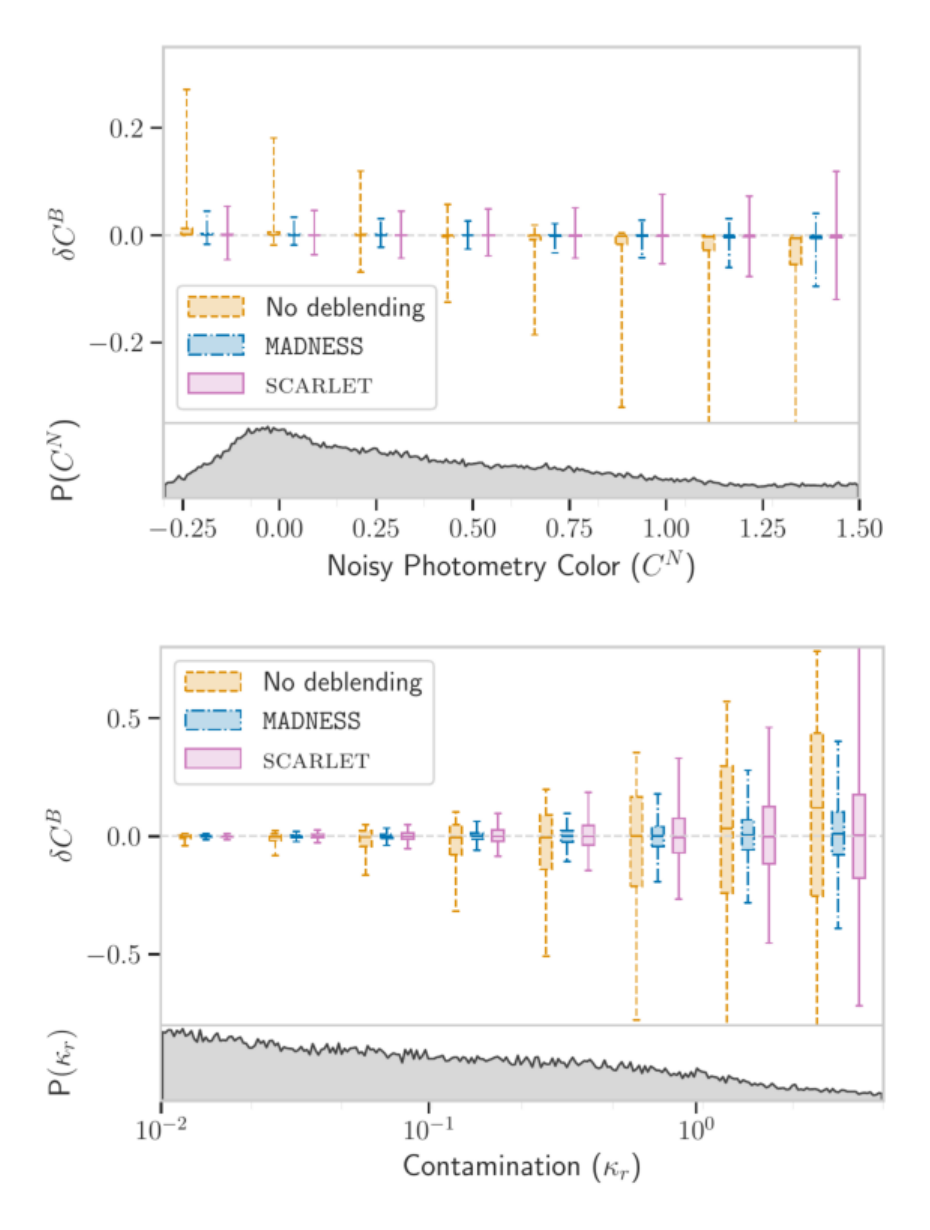}
         \label{fig: madness vs scarlet color}
     }

    \caption{Comparison of the \textit{g-r} color for galaxies in \TestFields for three cases: without deblending, deblending with \MADNESS and with \scarlet, as we vary the color (top panels) and contamination, $\kappa_r$, in the \textit{r}-band (bottom panels).
    Panel (a) shows the variation of true color residuals ($\delta C^T$) computed with the color from catalog ($C^T$) as the baseline, and panel (b) shows the distribution of blending error ($\delta C^B$) with colors from photometry on noisy isolated galaxies ($C^N$) as the baseline.
    For each panel, the top plot shows the color residuals where the boxes and whiskers in the plot indicate the ranges from $25 ^{\rm{th}}$ to $75 ^{\rm{th}}$ and $5 ^{\rm{th}}$ to $95 ^{\rm{th}}$ percentiles, respectively.
    The plot at the bottom of each panel shows the distribution of the metric on the horizontal axis.}
    \label{fig: color results}
\end{figure*}

In Section \ref{subsec: Aperture Photometry}, we saw that \MADNESS improved the flux reconstruction in the \textit{r}-band of LSST. 
However, we must also ensure the cross-calibration between bands is conserved, as these errors and biases may propagate into downstream analysis, into redshift estimation, for example.
In this section, we compare the \textit{g-r} color obtained with the flux in Section \ref{subsec: Aperture Photometry} for galaxies in \TestFields for three cases: without deblending, deblending with \MADNESS, and deblending with \scarlet.
For each of these three cases, we compute the error in measured color against two baselines: the true color from the catalog and the color obtained with photometry on noisy isolated galaxies.

\subsubsection{True residuals} 
\label{subsubsec: True Error}

For the baseline, we used the true color value obtained from the catalog ($C^T$). Figure \ref{fig: Photometry Bias} shows the variation of the true residual ($\delta C^T$) mathematically given by
    \begin{equation}
        \delta C^T = C_m - C^T,
    \end{equation}
    where $m\in\{\text{no deblending, \MADNESS, \scarlet}\}$ and correspondingly $C_m$ refers to the \textit{g-r} color measured on \TestFields in the three cases: without deblending, deblending with \MADNESS, and deblending with \scarlet. 
    We see in Figure \ref{fig: Photometry Bias} that even for low contamination, errors are induced by the noise in images, but they are symmetrical around zero so there is no bias. 
    As the contamination increases, the symmetry breaks and the uncertainty increases. However, both deblenders efficiently remove the bias and reduce the errors introduced by blending.
    
    The distribution of $\delta C^T$ shows a subtle skew as the contamination increases, and in different bins of true color, we observe a slight bias being introduced in the bins at both ends of the plot.
    This pattern of bias is expected because of the statistics of true galaxy color - i.e., a galaxy with a very low value for \textit{g-r} color (for example, $-0.5$) is most likely to be blended with a galaxy of higher color value, which will cause the observed value of \textit{g-r} color to increase and vice versa. 

    Although $\delta C^T$ represents the actual residuals in the color measurement, it includes the errors from both the effect of blending and the effect of noise, thus limiting our ability to compare the deblenders efficiently.

\subsubsection{Blending residuals} 

To remove the impact of noise in the measurement and evaluate the deblenders, we now compare $C_m$ against the color estimated by aperture-photometry of the galaxies in isolated noisy scenes ($C^N$).
The residuals from blending, $\delta C^B$, are then given as
\begin{equation}
    \delta C^B = C_m - C^N,
\end{equation}
where once again $m\in\{\text{no deblending, \MADNESS, \scarlet}\}$.
The distributions of $\delta C^B$ in Figure \ref{fig: madness vs scarlet color} back our conclusions from Section \ref{subsubsec: True Error} that blending leads to a bias in our measurement of the color. 
Moreover, with $\delta C^B$ we can directly compare the deblenders. 
We see in Figure \ref{fig: madness vs scarlet color} that although both \MADNESS  and \scarlet reduce the error and bias in color, \MADNESS reconstructs the \textit{g-r} color more accurately.

\subsection{Errors in morphology and shapes of the deblended galaxies}
\label{subsection: morphology}

\begin{figure}
\centering
\includegraphics[width=.5\textwidth]{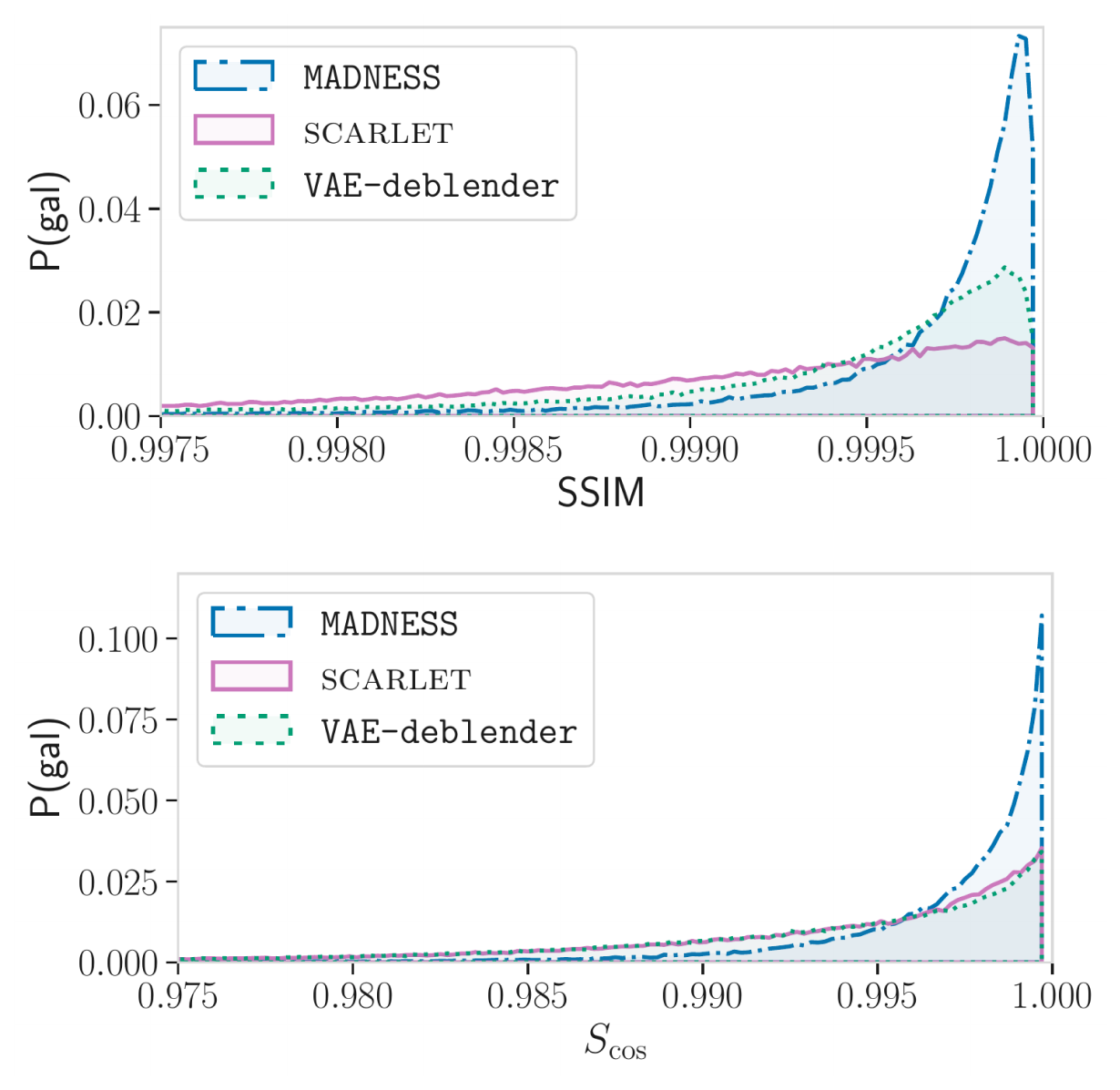}
\caption{Distribution of scores for morphology metrics: SSIM and cosine similarity ($S_{\text{cos}}$) for the deblended models of \MADNESS, \scarlet, and \VAEdeblender in the \textit{r}-band of galaxies from the \TestFields.
For both metrics, values closer to 1 imply a better performance.}
\label{fig: morphology results}
\end{figure}

\begin{figure}
     \centering
     \subfloat[]{
         \centering
         \includegraphics[width=.49\textwidth]{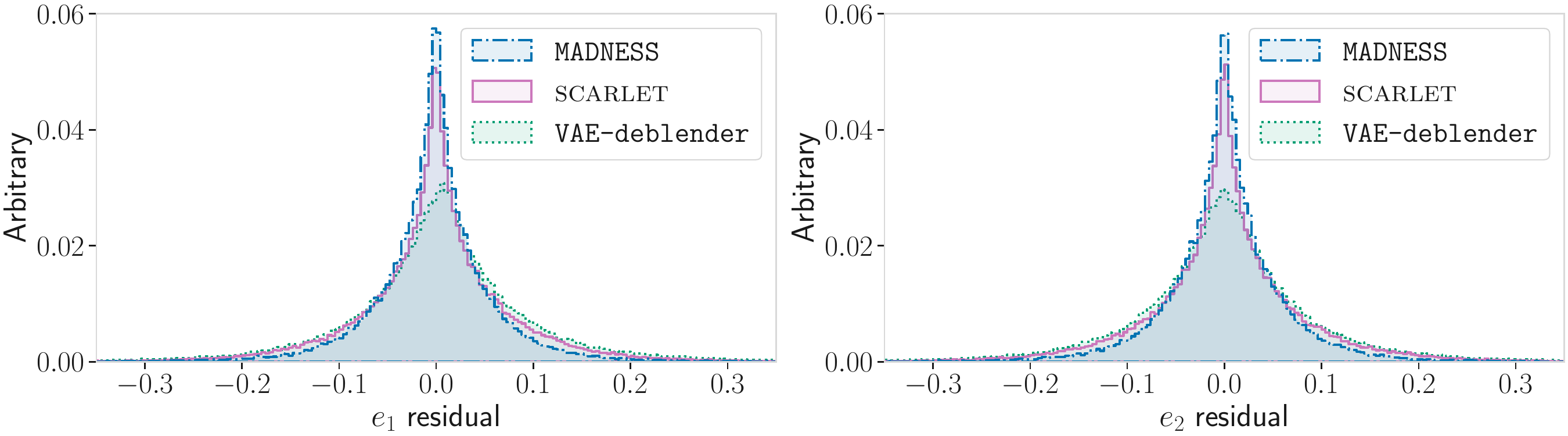}
         \label{fig: shapes uncut}
     }
     \hfill
     \subfloat[]{
         \centering
         \includegraphics[width=.49\textwidth]{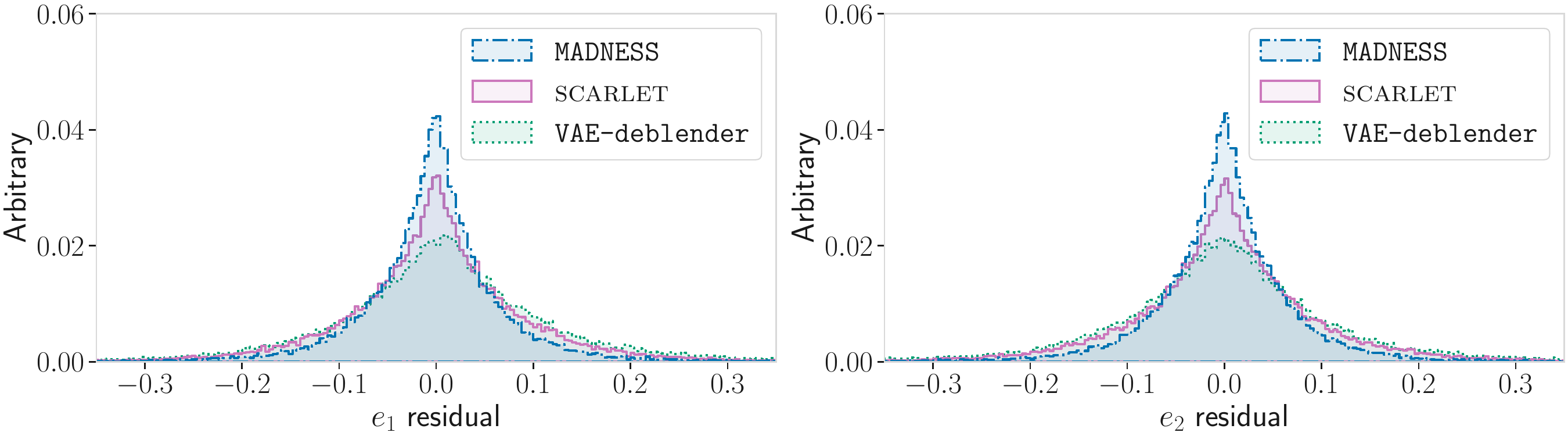}
         \label{fig: shapes cut}
     }

    \caption{Comparison of the ellipticities from second-order moments among \MADNESS,  \scarlet, and \VAEdeblender.
    The ellipticity residuals are shown for (a) entire \TestFields dataset and (b) slightly blended galaxies with $\kappa_r>0.01$, which represents an at least $1\%$ flux contribution from neighboring sources.}
    \label{fig: shapes}
\end{figure}

\begin{figure}
\centering
\includegraphics[width=.5\textwidth]{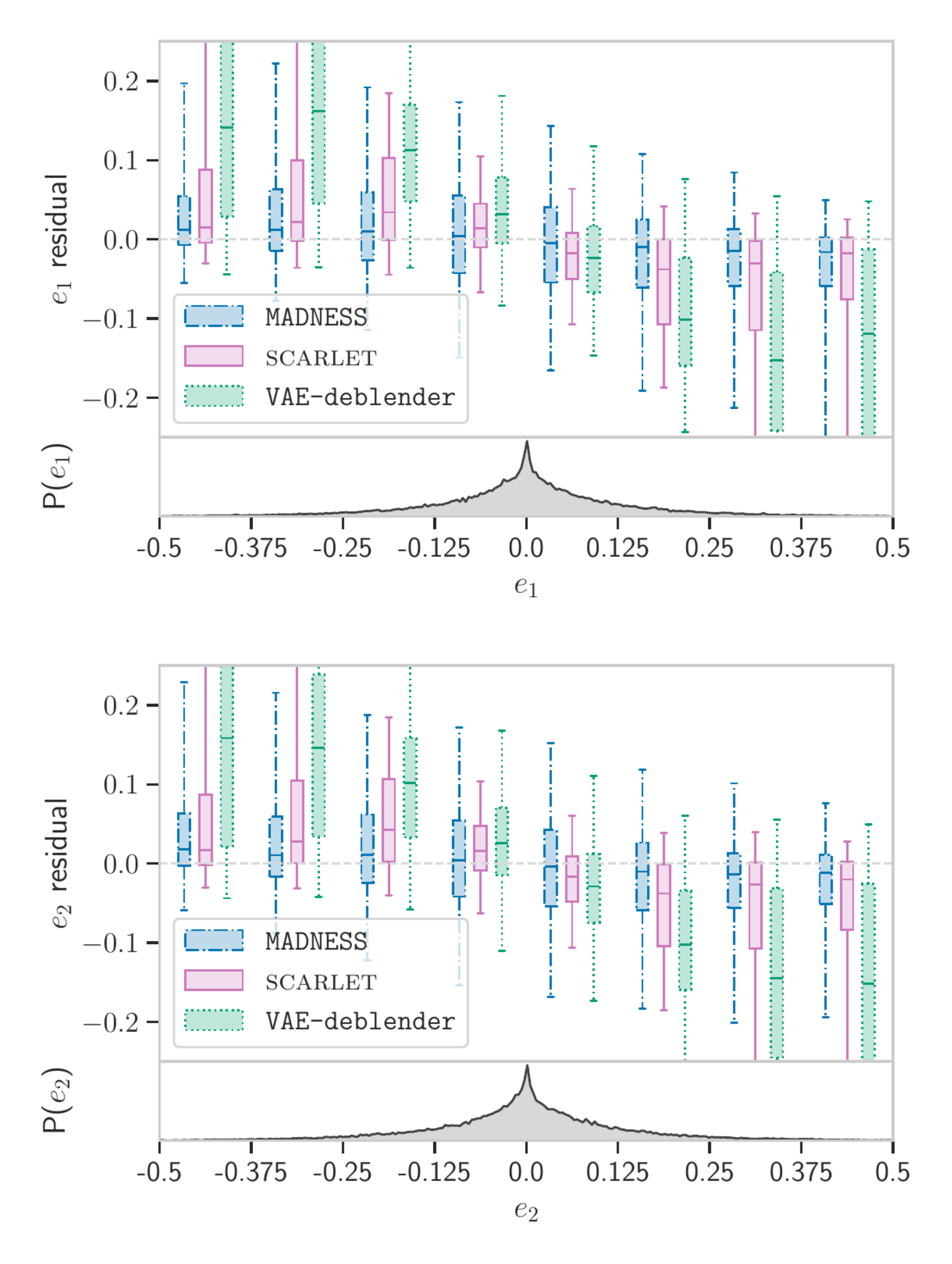}
\caption{Distribution of ellipticity residuals for the deblended models of \MADNESS, \scarlet, and \VAEdeblender in the \textit{r}-band of galaxies from the \TestFields
in different bins of $e_1$ (top) or $e_2$ (bottom). The boxes and whiskers in the plot indicate the ranges from $25 ^{\rm{th}}$ to $75 ^{\rm{th}}$ and $5 ^{\rm{th}}$ to $95 ^{\rm{th}}$ percentiles, respectively.}
\label{fig: shapes boxplot results}
\end{figure}

In this section, we evaluate the morphology of the reconstructed galaxies. 
Besides the SSIM metric, introduced in Section \ref{subsec: VAE}, we also use the cosine similarity $S_{\text{cos}}$, both of which give an idea of the similarity in structural information in the images.
Mathematically, $S_{\text{cos}}$ is given by

\begin{equation}
    S_{\text{cos}} =  \frac {\sum_{i=1}^N G_i P_i}{\sqrt{\sum_{i=1}^N G_i^2}\sqrt{\sum_{i=1}^N P_i^2}},
\end{equation}
where $G$ represents the simulated true pixel value for the galaxy, $P$ refers to the predicted galaxy model, the index $i$ refers to the $i ^{\rm{th}}$ element of the vector, and $N$ is the number of pixels. To compute this metric, we flatten the 2D image in each band into a 1D array; the metric can be interpreted as the cosine of the angle between these two 1D vectors.

Both SSIM and $S_{\text{cos}}$ metrics range from $-1$ (anti-correlation) to $+1$ (exact prediction).
To evaluate the deblenders, we computed the two metrics between the simulated galaxy and the deblended models separately in each band.
In Figure \ref{fig: morphology results}, we compare the SSIM and cosine similarity of the reconstructions in the \textit{r}-band. 
We observe that the performance of the \VAEdeblender is very close to that of \scarlet, and \MADNESS performs better than both \scarlet and the \VAEdeblender. To be more quantitative, with galaxies from \TestFields satisfying $\kappa_r \leq 5$, 
the average value for $S_{\text{cos}}$, for example, is $0.994$, $0.991$, and $0.987$ for \MADNESS, \scarlet, and \VAEdeblender, respectively.

To evaluate the shapes of the reconstructions, we used a definition of ellipticity $e$ that is commonly used in weak lensing analyses\footnote{See Part 3, Fig. 2 in \cite{schneider_gravitational_2006}, for example.}: 
\begin{equation}
    e = e_1 + i e_2 = \frac{Q_{xx} - Q_{yy} + 2iQ_{xy}}{Q_{xx} + Q_{yy}},
\end{equation}
where $Q_{ij}$ are the second moments of galaxy images.

We compare the ellipticity residuals defined as the difference between the ellipticities $e_1$ or $e_2$ of the deblended models and the values obtained from isolated noiseless galaxy images. 
The distribution of ellipticity residuals in Figure \ref{fig: shapes uncut} shows that for the entire \TestFields dataset \MADNESS performs better than both \scarlet and the \VAEdeblender. 
For example, the standard deviation of the $e_1$ residuals is $0.069$, $0.084$, and $0.104$ for \MADNESS, \scarlet, and \VAEdeblender, respectively.
Figure \ref{fig: shapes cut} shows the distribution of residuals upon selecting only blended galaxies with contamination $\kappa_\textit{r} \geq 0.01$.
The standard deviation for $e_1$ residuals with the contamination cut is $0.087$ for \MADNESS and $0.103$ for \scarlet. 
The relatively better performance of \MADNESS compared to \scarlet is in agreement with our results from Section \ref{subsec: Aperture Photometry}.

In Figure \ref{fig: shapes boxplot results} we show from the distribution of ellipticity residuals in bins of $e_1$ and $e_2$ that although all deblenders are biased towards predicting rounder galaxies, i.e., smaller absolute values for the ellipticities, \MADNESS generates more consistent results across the different bins of $e_1$ and $e_2$. 
To reduce these biases further for both \MADNESS and the \VAEdeblender, appropriate cuts can be applied to the training dataset to train the models with more elliptical galaxies.

\section{Towards real data}
\label{sec: towards real data}

Before we apply \MADNESS to real data, a few important steps require consideration. \MADNESS must be combined with a detection algorithm, as it does not include a detection step. 
In our work, we restricted ourselves to the ground-truth positions to evaluate the efficiency of the deblender, but in reality, a significant fraction of these blends are likely to go unrecognized, and the galaxies may not be perfectly centered.
Additionally, it might be necessary to retain the networks with lower S/N objects to handle false detections.
Moreover, multi-band processing will also add subtleties such as coaddition and World Coordinate Systems (WCS), which may vary across bands.
Further work is required to systematically evaluate the biases introduced during the deblending process when \MADNESS is combined with a detection algorithm. 
In fact, the final value of log-prior in Equation \ref{equation: final deblending solution} may provide valuable insight to flag potentially unrecognized blends, and space-based surveys can be useful for obtaining prior detections, which adds to the motivation for the joint-processing of data.

In this work, we have only considered bulge plus disk models and for each galaxy. Except for the magnitudes and PSF, all galaxy parameters related to shapes and sizes are assumed to be constant across all bands. This assumption is not necessarily the case for real data \citep{2022MNRAS.516..942C, 2022A&A...664A..92H}, and this warrants further follow-up studies with more realistic simulations consisting of complex galaxy morphologies.
Additionally, the algorithm needs to be flexible enough to handle stars properly. 
This can be done either by including stars in the training dataset of the VAE itself so that the generative model can simulate both stars and galaxies or by having a hierarchical framework that includes a dedicated star galaxy separation algorithm \cite[see][]{hansen2022scalable}. 

Finally, our work can also be extended into a hierarchical Bayesian framework to incorporate PSF information extracted by other science pipelines. 
Following the same ideas as \citet{10.1093/mnras/stab1214}, \MADNESS can be trained to predict the PSF-deconvolved galaxies and reconvolved with externally obtained PSF before computing the likelihood. 
Furthermore, the lower level of blending in space-based telescopes due to smaller PSFs implies that joint processing data can significantly help to deblend ground-based observations.
\citet{10.1093/mnras/staa3062} already showed that VAEs can be used for multi-instrument processing by concatenating data as different channels to the input layer of the VAE.
In particular, the authors show that the results of deblending were significantly improved when combining data from Euclid with Rubin.
Since \MADNESS also makes use of a VAE, we expect this to generalize and hope to obtain a significant boost by multi-instrument processing.

\section{Conclusions}
\label{sec:Conclusions}

Blending is expected to be a major source of systematic errors in the next generation of cosmological surveys.
The complexity of the problem, given the wide range of morphologies and colors of astrophysical objects, makes it an ideal domain for the application of ML techniques that have already proven to be extremely efficient for image generation and processing.
In this context, we developed a new deblending algorithm called \MADNESS, which combines two generative models, namely the VAE and normalizing flow, in a way so that we can obtain the MAP solution for the inverse problem of deblending.

We trained and tested our algorithm with galaxies expressed as a sum of bulge and disk models taken from the \CatSim catalog. The galaxies were simulated for a 10-year LSST survey using \galsim with the help of \btk \citep{Mendoza_2024BTK}, which allows one to easily control the number and position of galaxies.
We first trained a $\beta-$VAE as a generative model for isolated galaxies and then trained a normalizing flow to learn the underlying aggregate posterior distribution in the VAE latent space.
The normalizing flow gives access to the log-prior term in the VAE latent space, and the decoder provides the mapping from the latent space to the image space, thus enabling the computation of the log-likelihood term needed to obtain the MAP solution.
Combining the two models, we perform a gradient descent in the VAE latent space to obtain a MAP estimate.
Typically, deep learning-based deblenders require training data to be representative of real data. 
While this is also true for \MADNESS, the optimization step makes our algorithm less susceptible to this than traditional feed-forward networks and at the same time, it makes our algorithm more interpretable.

To enable faster convergence of our algorithm, we retrained the encoder to obtain an initial position to start the optimization. 
We call this new encoder the \VAEdeblender because the encoder is trained to find the latent space representation of the central galaxy in a blended postage stamp, thereby deblending it from neighboring sources, similar to the network proposed by \citet{10.1093/mnras/staa3062}. 
Although the \VAEdeblender is very efficient in itself, the results thus obtained are subject to the black-box nature of the algorithm and lack explicit flux optimization. 
In our algorithm, we take the latent space representations obtained from the \VAEdeblender as the initial point and perform a gradient descent in the VAE latent space to obtain the MAP solution.

We tested our algorithm for flux reconstruction with aperture-photometry, the \textit{g-r} color of the deblended models, and morphology using SSIM and the pixel cosine similarity as metrics. 
As a sanity check, we compared our results to the initial deblending solutions of the \VAEdeblender, and observed a significant improvement for all metrics in the final MAP solution. 
We also compared \MADNESS to the state-of-the-art deblender \scarlet.
Looking at the results in the \textit{r}-bandpass filter, we observed that the \VAEdeblender was already able to closely match the performance of \scarlet. 
However, when we look at the results from other filters in Appendix \ref{section: appendix photometry results} and \ref{section: appendix morphology results} of LSST, we notice a significant drop in performance of both \MADNESS and the \VAEdeblender compared to \scarlet in the \textit{u}-band, especially in terms of morphology.
This is because \scarlet predicts a joint morphology for all bands and the high-fidelity bandpass filters (such as \textit{r}, \textit{i}, etc.) allow better reconstructions.

In the case of \MADNESS and the \VAEdeblender, although the VAE learns correlations between different bands, the predicted morphology can still vary between the different channels of VAE output.
When obtaining the MAP estimate, \MADNESS tries to minimize the data likelihood, so we expect the overall flux to be consistent even in low-S/N bands. 
However, reconstructing the morphology is more difficult as the edges of the galaxies will have even lower S/N than the central brighter regions. 

Our algorithm can be easily expanded to a hierarchical framework to incorporate physical information, such as PSF, along with noise.
Furthermore, it can also be extended into a multi-instrument approach by concatenating data as new channels to the VAE input.
\citet{10.1093/mnras/staa3062} already showed that doing so can improve the deblending results, and we expect them to generalize.
Such algorithms that are capable of combining data at the pixel level are likely to play a key role in the upcoming golden era of observational cosmology, where we will have data from several stage IV ground- and space-based surveys.

\begin{acknowledgements}

This paper has undergone internal review in the LSST Dark Energy Science Collaboration. The authors would like to kindly thank the internal reviewers, Patricia Burchat and Cyrille Doux. Their comments, feedback, and suggestions helped us greatly improve the paper. We also would like to thank François Lanusse, Ismael Mendoza,
Peter Melchior, Fred Moolekamp, James J. Buchanan and 
Tianqing Zhang for comments and discussions throughout this work.
      
B.B. led the project, wrote the code, trained the neural networks, performed the analysis, and wrote the initial draft of the paper. 
E.A., A.B., A.G., J.L., and C.R. supervised the project, reviewed the code, and provided feedback on the draft.

This project has received funding from the European Union’s Horizon 2020 research and innovation program under the Marie Skłodowska-Curie grant agreement No 945304 – COFUND AI4theSciences hosted by PSL University, and Agence Nationale de la Recherche grant ANR-19-CE23-0024 — AstroDeep. We gratefully acknowledge support from the CNRS/IN2P3 Computing Center (Lyon - France) for providing computing and data-processing resources needed for this work.

The DESC acknowledges ongoing support from the Institut National de 
Physique Nucl\'eaire et de Physique des Particules in France; the 
Science \& Technology Facilities Council in the United Kingdom; and the 
Department of Energy and the LSST Discovery Alliance
in the United States.  DESC uses resources of the IN2P3 
Computing Center (CC-IN2P3--Lyon/Villeurbanne - France) funded by the 
Centre National de la Recherche Scientifique; the National Energy 
Research Scientific Computing Center, a DOE Office of Science User 
Facility supported by the Office of Science of the U.S.\ Department of
Energy under Contract No.\ DE-AC02-05CH11231; STFC DiRAC HPC Facilities, 
funded by UK BEIS National E-infrastructure capital grants; and the UK 
particle physics grid, supported by the GridPP Collaboration.  This 
work was performed in part under DOE Contract DE-AC02-76SF00515.

The authors acknowledge the use of the following libraries: \texttt{numpy}, \texttt{astropy}, \TensorFlow, \TFP, 
\scarlet, \BlendingToolkit, \galsim, \SurveyCodex, \SourceExtractor, and \texttt{matplotlib}.

\end{acknowledgements}

\bibliographystyle{aa} 
\bibliography{refs}

\begin{appendix}

\section{S/N computation}
\label{section: S/N computation}

Following the LSST documentation\footnote{\url{https://smtn-002.lsst.io/}}, S/N is defined as
\begin{equation}
    \text{S/N} = \frac{C}{\sqrt{C + (\frac{B}{g} + \sigma^2_{\text{instr}})n_{\text{eff}}}} ,
\end{equation}
where $C$ is the total source counts, $B$ is sky background counts per pixel, $\sigma_{\text{instr}}$ is the instrumental noise per pixel (all in ADU), $g$ is the gain, and $n_{\text{eff}}$ refers to the source footprint. We discard the $\sigma_{\text{instr}}$ term as our simulations do not take it into account. We used \SurveyCodex to obtain the $B$ and $C$ parameters for each galaxy and set the value of gain to $1$ for consistency with our \btk simulations.
Finally, we evaluate $n_{\text{eff}}$ as
\begin{equation}
    n_{\text{eff}} = 2.266(\text{FWHM}_{\text{eff}}/P)^2 ,
\end{equation}
where $\text{FWHM}_{\text{eff}}$ is the effective full width at half maximum and $P$ refers to the pixel scale, which for LSST is $0.2$\arcsec per pixel.
We compute the $\text{FWHM}_{\text{eff}}$ as a convolution of PSF FWHM and galaxy FWHM ($\text{FHWM}_{\text{gal}}$) given by
\begin{equation}
   \text{FHWM}_{\text{gal}} = 2\sqrt{a_{\text{hlr}}b_{\text{hlr}}},
\end{equation}
where $a_{\text{hlr}}$ and $b_{\text{hlr}}$ are the semi-major and semi-minor axes of either the bulge or the disk, whichever is greater.

\section{Architectures and training of networks}
\label{section: Network Architectures and Training}

\subsection{VAE}
\label{subsec: VAE HyperParameters}
\begin{figure*}
     \centering
    \includegraphics[width=\textwidth]{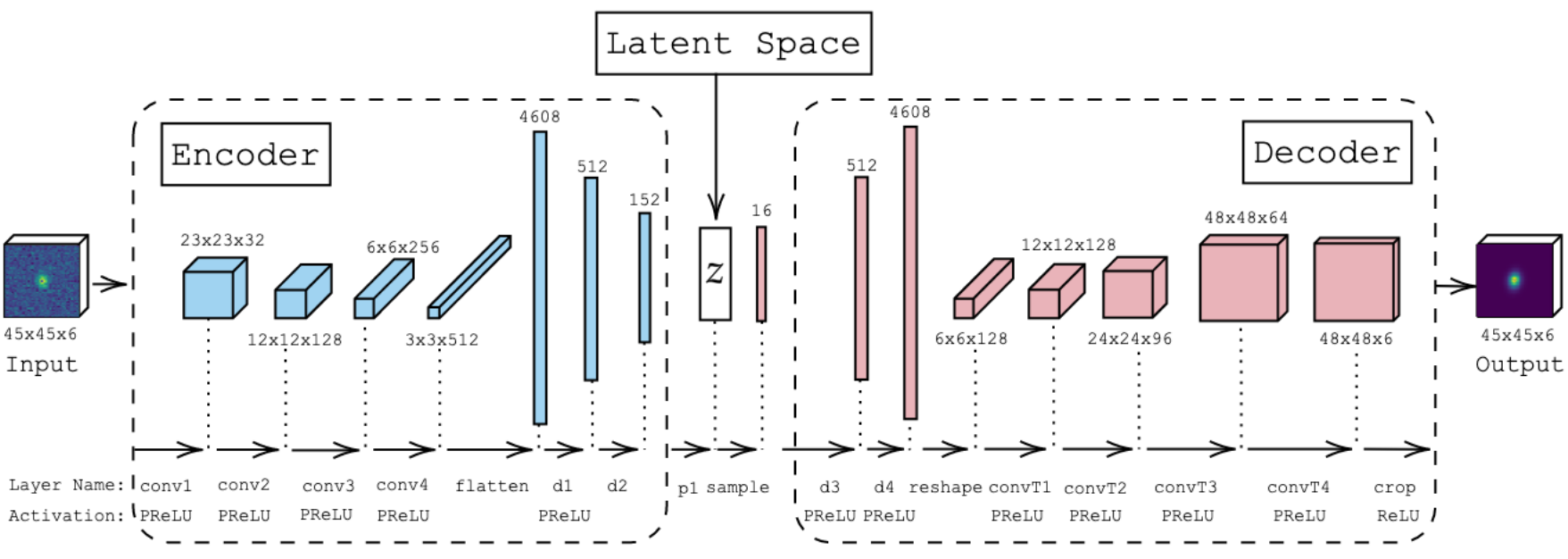}
    \caption{Variational autoencoder architecture. The \texttt{conv1}, \texttt{conv2}, \texttt{conv3}, and \texttt{conv4} layers refer to the four convolutional layers of the encoder, all of which use a kernel size of 5 and a stride of 2. Following the convolutional layers are the dense layers \texttt{d1} and \texttt{d2} that predict the parameters of the latent space distribution. The probabilistic layer \texttt{p1} uses the output of layer \texttt{d2} to construct the latent space $z$, which is a multivariate normal distribution of 16 dimensions. Finally, the decoder takes as input a sample from the latent space distribution and applies a set of dense layers \texttt{d3} and \texttt{d4}, followed by transposed convolutional layers (\texttt{convT1}, \texttt{convT2}, \texttt{convT3}, and \texttt{convT4}) also with kernel size 5 and stride 2; Finally, the output of \texttt{convT4} is cropped to match the size of the input.}
    \label{fig: VAE diag}
\end{figure*}
In our implementation, we used the \TensorFlow\footnote{\url{https://www.tensorflow.org/}} framework and \TFP\footnote{\url{https://www.tensorflow.org/probability}} for the probabilistic layers.
As shown in Figure \ref{fig: VAE diag}, the encoder has 4 convolutional layers with 64, 128, 256, and 128 filters. 
Each layer uses a convolution kernel of size 5 and a stride of 2. 
These layers are followed by 2 dense layers that predict the latent space distribution. 
Each layer in the encoder is followed by \texttt{PReLU} activation functions except the final dense layer, which is left without an activation function as it predicts the parameters for a multivariate normal distribution layer that describes the $16$-dimensional latent space distribution.
The latent space is followed by the decoder, which also has two dense layers and 4 transposed convolutional layers with 128, 96, 64, and 6  filters. 
The first 3 layers use a kernel of size 5 and stride 2, while the final layer has a kernel of size 3 and stride 1. 
We used the \texttt{PReLU} activation function in all the layers except the final layer, where we applied a \texttt{ReLU} activation function to prevent negative predictions of flux.
Finally, the output is cropped to match the size of the input.
The encoder and decoder have around $6.7$M and $3.5\rm{M}$ trainable parameters, respectively.
The architecture of the encoder is chosen to be more complex because it is later trained as a deblender to obtain the initial point for gradient descent, as described in Section \ref{subsection: deblender}.

During training, we optimized the loss function in Equation \ref{equation: ELBO warm-up}, where $\beta$, the KL divergence coefficient, is set to $0.01$.
By using $\beta<1$, we can improve the reconstruction fidelity, but this also implies that the learned aggregate posterior differs from the prior.
An alternative approach could be to remove the KL term entirely. However, the lack of regularization will lead to a more complex latent space distribution, as there is no incentive for the VAE to follow the ELBO-prior.
The added complexity of the latent space would make it more difficult to perform a gradient descent in the VAE latent space for the MAP estimate.
After testing the $\beta-$VAE for several values of the hyperparameter $\beta$, we chose the value of $0.01$ as no significant improvement was observed by lowering the value further.

The coefficient $\alpha$ is initialized at $1$ and is decreased by $0.02$ every epoch until it reaches a value of $0$ after $50$ epochs. 
The SSIM is computed on normalized images where each band of input and output is divided by the maximum pixel value in that band, so all values are in the range $[0, 1]$.
Since the $\ssim$ is equally weighted among all the bands, the network is forced to learn the features in all the filters from the very beginning. 

We used the Adam optimizer \citep{kingma2017adam} with an initial learning rate of $10^{-5}$, along with a learning rate scheduler that decreases the learning rate by a factor of $2.5$ every $40$ epochs. 
Additionally, since we performed only a linear normalization of our data and some galaxies are extremely bright compared to others, we introduce a clip value of $0.1$ on the gradients obtained during the optimization to stabilize the training of our network.
After the initial warm-up phase is complete, we dropped the SSIM term from the loss function and restarted the training with a learning rate of $10^{-5}$ for a maximum of $200$ epochs with an early stopping patience of $20$ so that the network stops training if the loss has not decreased over $20$ iterations. 
Even during this training phase, we keep the same learning rate scheduler and clip value.

\subsection{Normalizing flows}
\label{subsec: normalizing flow HyperParameters}

\begin{figure}
\centering
\includegraphics[width=.5\textwidth]{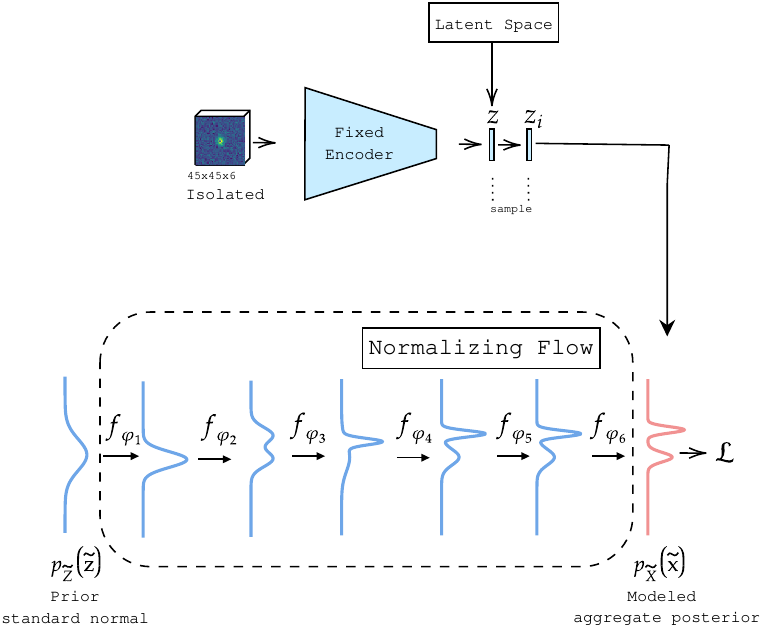}
\caption{Schematic of the training of the normalizing flow. The encoder is the pre-trained network illustrated in Figure \ref{fig: VAE diag} and the normalizing flow consists of 6 bijective transformations $f_{\varphi_i}$ where the $i^{th}$ transformation is parameterized by $\varphi_i$, which represents the weights of the MAF layer containing (32, 32) units with \texttt{tanh} activations followed by a permutation layer. Additionally, a \texttt{batchnorm} layer is added as the first layer within $f_{\varphi_1}$.}
\label{fig: normalizing flow}
\end{figure}

We used \TFP to create bijective autoregressive networks with 2 hidden layers containing 32 units each followed by \texttt{tanh} activation functions, to be used within MAF layers.
After an initial \texttt{batchnorm} layer, we use 6 sets of MAF followed by permutation and batchnorm layers, resulting in a model with a total of \num{15936} trainable parameters (see Figure \ref{fig: normalizing flow}).
We used the Adam optimizer and started with a learning rate of $10^{-4}$, clip value of $0.01$, and the same setting for the learning rate scheduler and patience as the VAE, for a maximum of $200$ epochs.

\subsection{\VAEdeblender}
\label{subsec: VAE-Deblender HyperParameters}

We ran this training for a maximum of $200$ epochs with the Adam optimizer and a learning rate of $10^{-5}$. Other parameters such as the clip value, learning rate scheduler, and patience were kept the same as the original VAE.

\section{Comparison between \textit{u}- and \textit{r}-band}
\label{Comparison between u and r-band}

Since the training of the VAE uses the likelihood obtained from pixel-wise predictions, the quality of reconstruction depends upon the S/N of the galaxies. 
In Figure \ref{Comparison between u and r-band}, we show a comparison between the quality of simulations of our trained model in the \textit{u} and \textit{r} bands for the same set of galaxies. 
This comparison demonstrates that the model is able to better learn the features in the higher S/N band compared to the lower S/N one. So we expect poor deblending reconstructions in the \textit{u}-band as shown in Appendix \ref{section: appendix photometry results} and \ref{section: appendix morphology results}.

\begin{figure}
     \centering
     \subfloat[]{
         \centering
         \includegraphics[width=.49\textwidth]{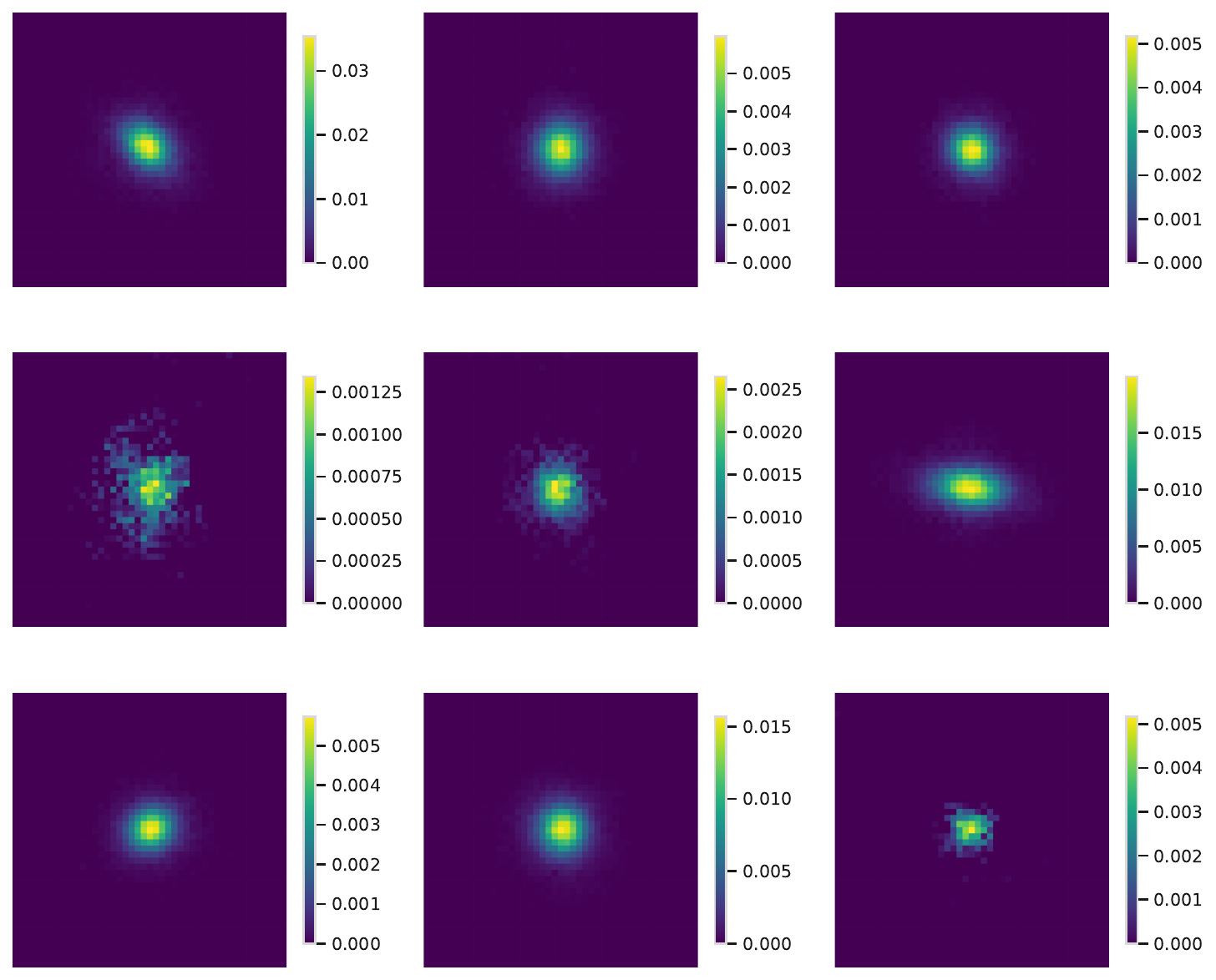}
    }
     \hfill
     \subfloat[]{
         \centering
         \includegraphics[width=.49\textwidth]{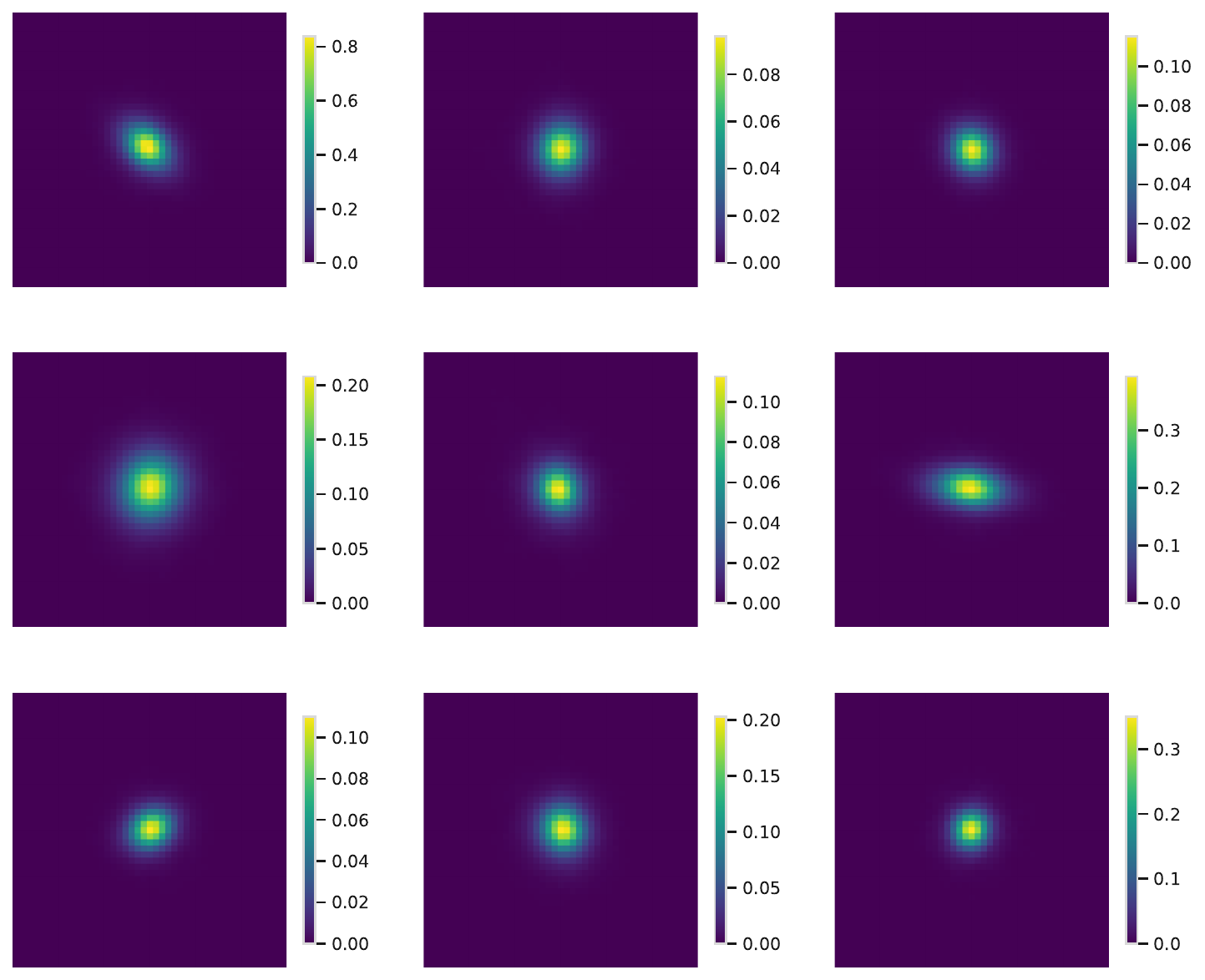}
     }

    \caption[Comparison between \textit{r} and \textit{u} band simulations]{Comparison of simulations for the same galaxies between the  \textit{u}-band shown in panel (a) and \textit{r}-band shown in panel (b). The color bar in each image shows the normalized electron counts in pixels.}
    \label{fig: generative model simulations r - u comparison}
\end{figure}

\section{\scarlet configuration}
\label{section: appendix scarlet config}

Using the \scarlet release \texttt{btk-v1}, all the galaxies were modeled as per standard recommendations in \scarlet documentation\footnote{\url{https://pmelchior.github.io/scarlet/}}.
The sources are initialized using the \texttt{scarlet.initialization.init\_all\_sources()} with all parameters set to default except for the \texttt{max\_components}, which is set to $2$ so that the bulge and disk components can be properly modeled, provided that the S/N is large enough.
The \texttt{min\_snr} parameter corresponding to this S/N value is set to the recommended default value of $50$.

On each field, the optimization was run for a maximum of \num{200} iterations, and following the tutorials for \scarlet, the relative error parameter determining the convergence was set to $10^{-6}$.

\section{Metrics}
\label{section: appendix metrics}

\begin{figure*}
\centering
\includegraphics[width=.98\textwidth]{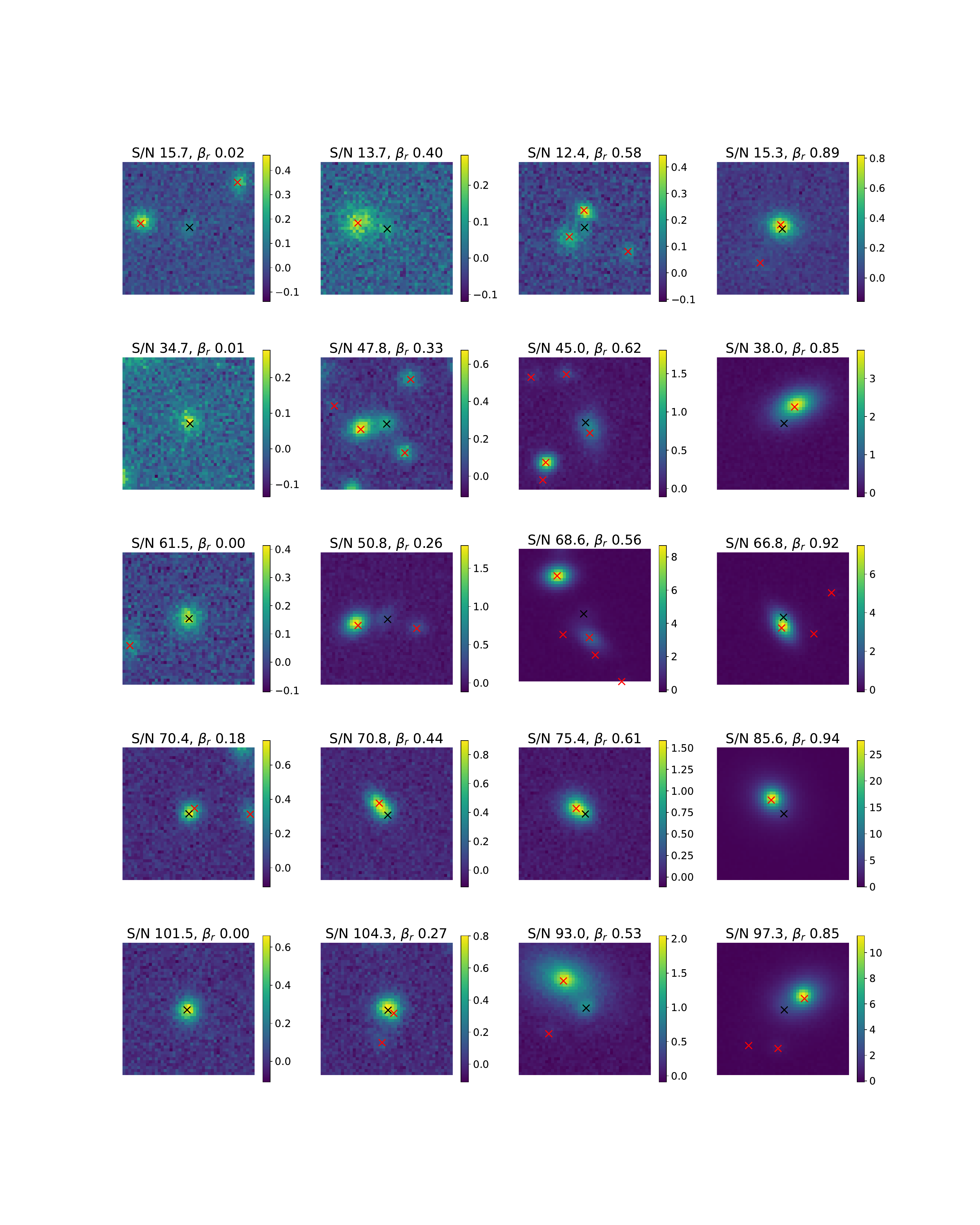}
\caption{Normalized images of galaxies in \textit{r}-band for different levels of blendedness $\beta_r$ of the central galaxy (centers marked with black $\times$), in the presence of neighboring galaxies (centers marked with red $\times$). From left to right, the value of blendedness increases, and from top to bottom, the S/N of the central galaxy increases. The color bar for each plot shows the normalized electron counts in the pixels.}
\label{fig: blendedness variation}
\end{figure*}

The blendedness ($\beta$) metric defined in Section \ref{subsec: Metrics} lies between 0 and 1, 0 for isolated galaxies, and close to 1 for severely blended ones. 
Figure \ref{fig: blendedness variation} gives a visual representation of the variation of blendedness for different values of S/N in the \textit{r}-band.

\begin{figure*}
\centering
\includegraphics[width=.98\textwidth]{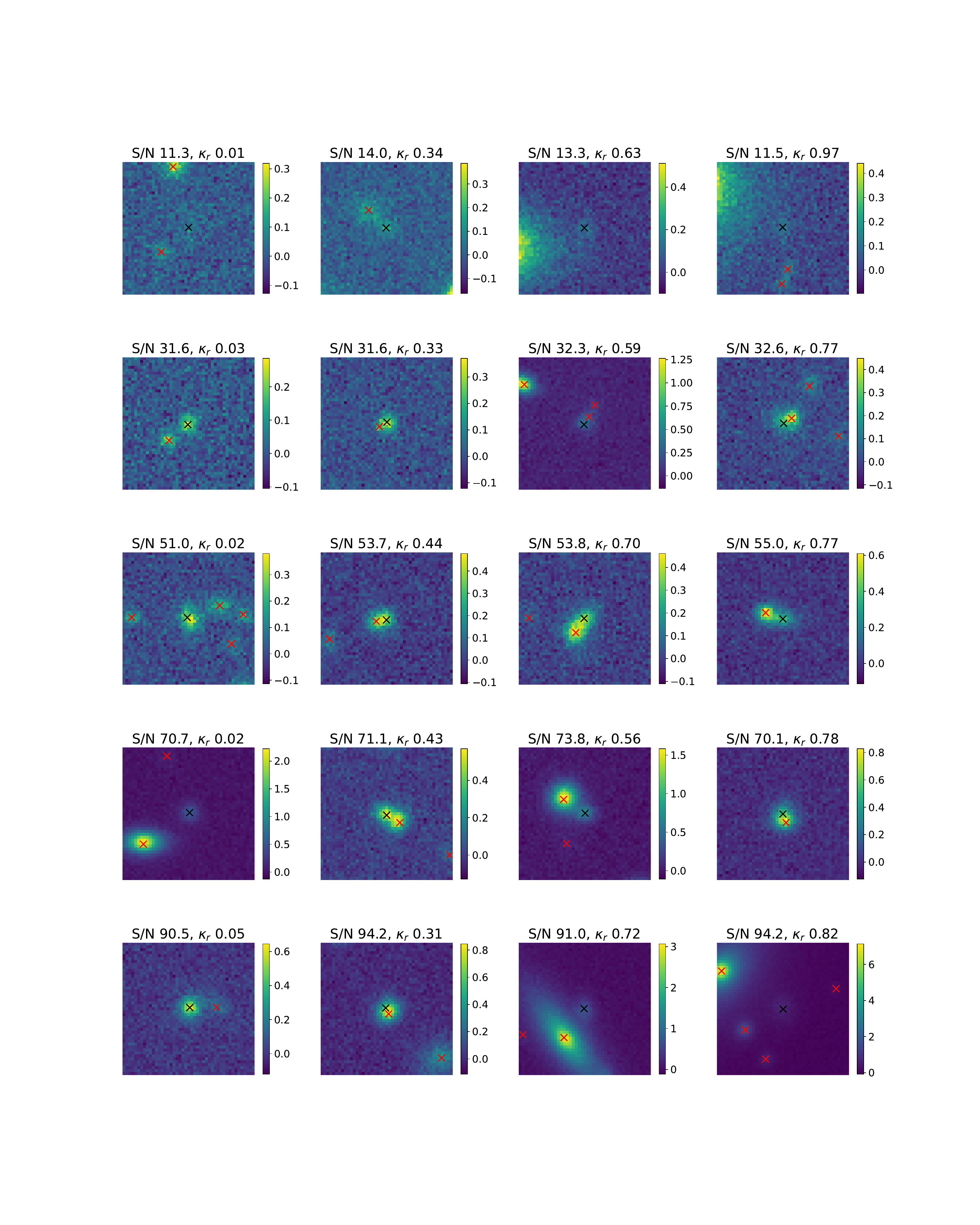}
\caption{Normalized images of galaxies in \textit{r}-band for different levels of contamination $\kappa_r$ of the central galaxy (centers marked with black $\times$), in the presence of neighboring galaxies (centers marked with red $\times$). 
From left to right, the value of contamination increases, and from top to bottom, the S/N of the central galaxy increases. The color bar for each plot shows the normalized electron counts in the pixels.}
\label{fig: contamination variation}
\end{figure*}

The contamination ($\kappa$) defined in \ref{subsec: Metrics} also quantifies the effect of blending. 
For isolated galaxies, the value of $\kappa$ is $0$ and increases according to the contribution to the flux from neighboring galaxies within the flux radius of aperture photometry.
Figure \ref{fig: contamination variation} gives a visual representation of how $\kappa$ varies in the \textit{r}-band for blended galaxies at different S/Ns.

\section{Photometry across bands}
\label{section: appendix photometry results}

\begin{figure*}
     \centering
     \subfloat[]{
         \centering
         \includegraphics[width=\textwidth]{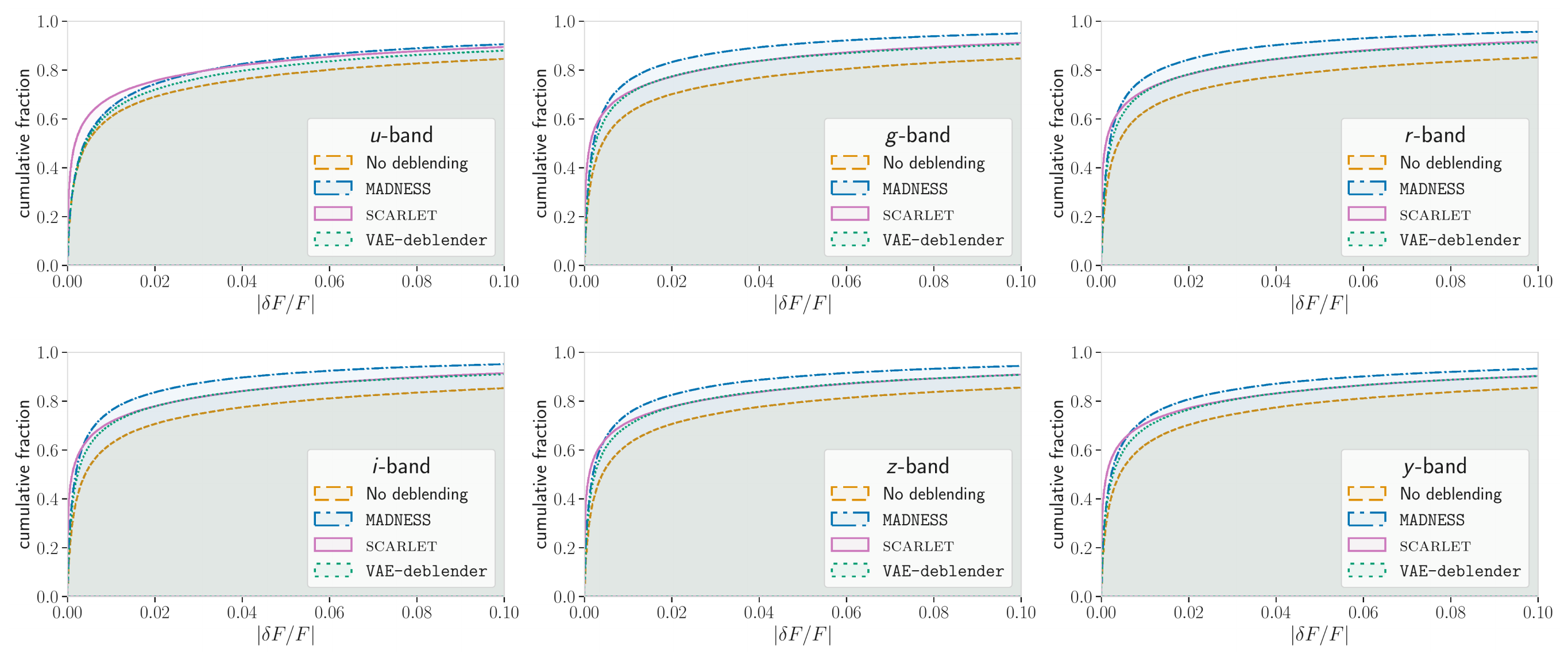}
         \label{fig: cumulative_distib_phot_err}
     }
     \hfill
     \subfloat[]{
         \centering
         \includegraphics[width=\textwidth]{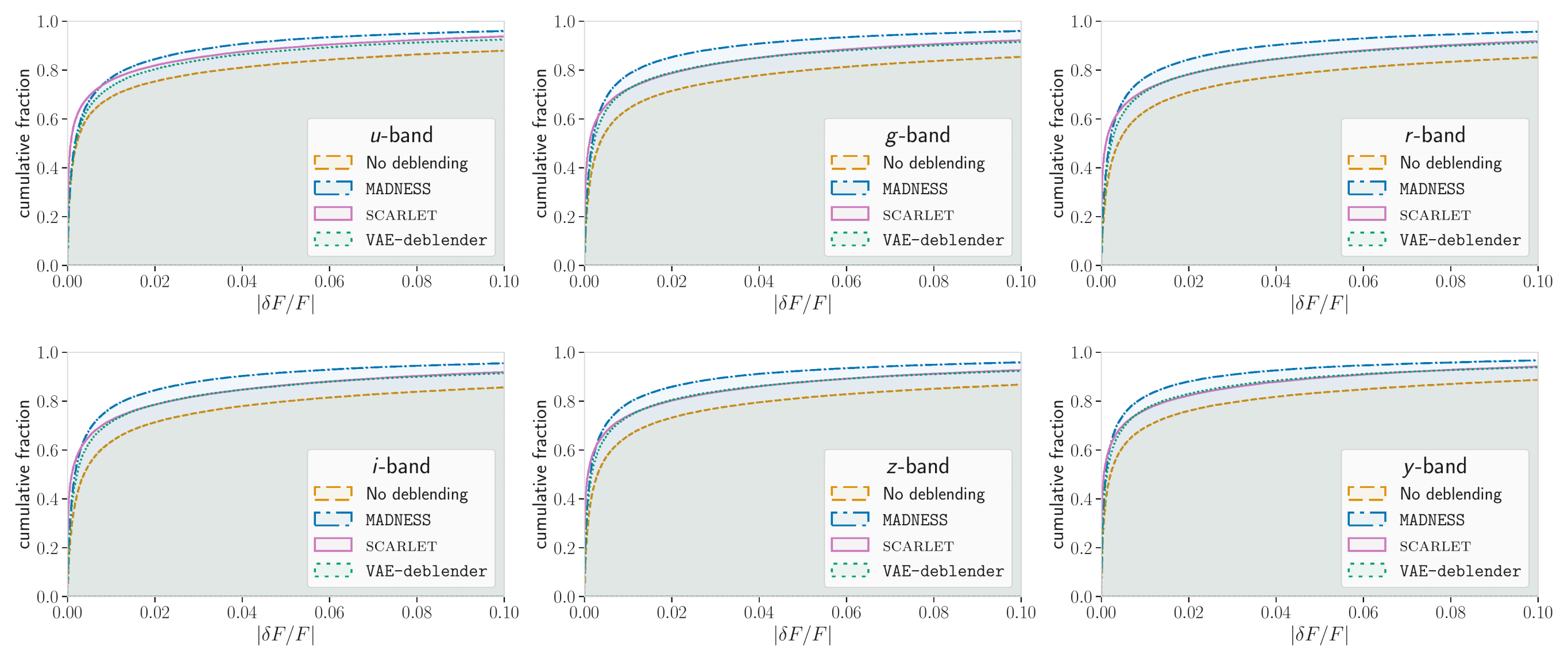}
        \label{fig: cumulative_distib_phot_err_masked}
     }

    \caption{Cumulative distribution of the absolute value of the relative flux residuals $|\delta F/ F|$ before deblending and after deblending with \MADNESS, \scarlet, and \VAEdeblender, for each LSST band. 
    The results are presented for two cases: (a) all galaxies in \TestFields and (b) galaxies in \TestFields with $\text{S/N}_{\text{blend}}\geq5$.}
    \label{fig: high-density photometry error}
\end{figure*}

In Section \ref{subsec: Aperture Photometry}, we showed that \MADNESS outperformed the flux reconstructions of \scarlet in the \textit{r}-band. In Figure \ref{fig: high-density photometry error}, compare the flux reconstructions among \MADNESS, \scarlet, and \VAEdeblender for all the bands of LSST. 

Figure \ref{fig: cumulative_distib_phot_err} shows the results for all the galaxies in the \TestFields dataset and we observe that the performance of \MADNESS and \VAEdeblender drops significantly \textit{u}-band, compared to that of \scarlet.
To see if this is due to the low S/N of galaxies in the \textit{u}-band, we redo the plot with only high S/N galaxies in each of the bands. 
To simplify the S/N computation in each band, we can define S/N\textsubscript{blend} for each band in the blended configuration as:
\begin{equation}
    \text{S/N}_{\text{blend}} = \texttt{flux}/\texttt{fluxerr};
\end{equation}
where \texttt{flux} and \texttt{fluxerr} are obtained from aperture-photometry with the \textsc{SourceExtractor} library.
For each band, after discarding galaxies with S/N\textsubscript{blend}$\leq5$ see in Figure \ref{fig: cumulative_distib_phot_err_masked} that now the performance of \MADNESS is comparable to \scarlet even in the \textit{u}-band. 
Therefore, we can conclude that the low S/N in \textit{u}-band leads to poor flux reconstructions.

\section{SSIM across bands}
\label{section: appendix morphology results}

\begin{figure*}
     \centering
     \subfloat[]{
         \centering
         \includegraphics[width=\textwidth]{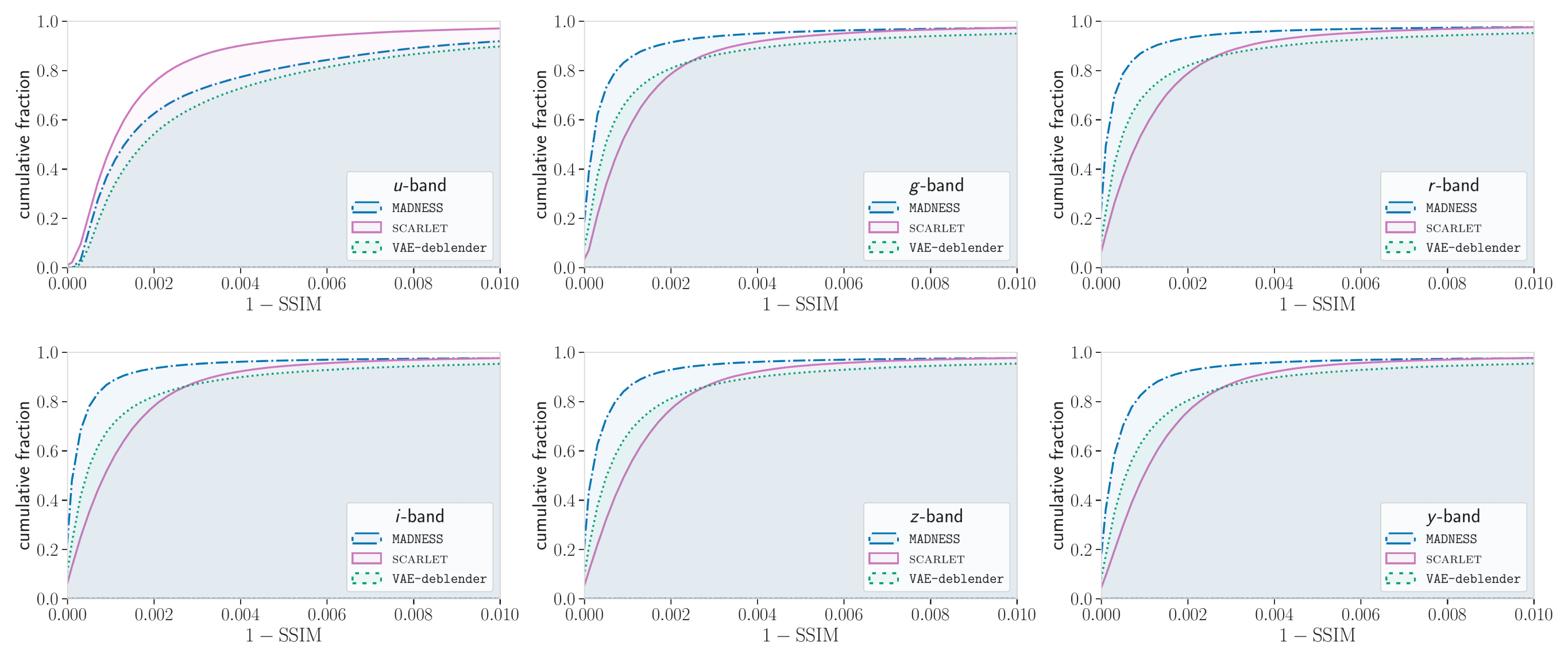}
         \label{fig: cosine similarity all bands}
     }
     \hfill
     \subfloat[]{
         \centering
         \includegraphics[width=\textwidth]{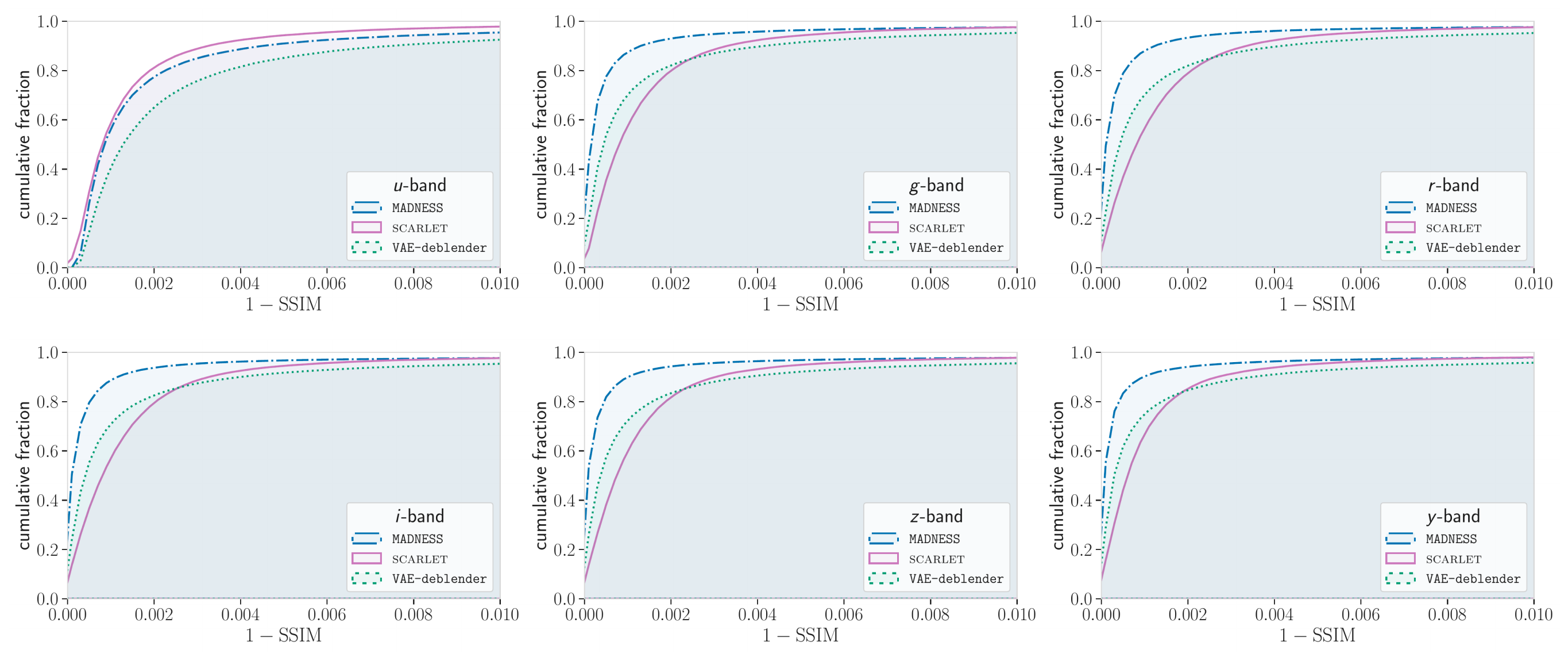}
         \label{fig: cosine similarity all bands masked}
     }

    \caption{
    Comparison of the structural similarity index measure (SSIM) after deblending with \MADNESS, \scarlet, and \VAEdeblender, for each LSST band. 
    The results are presented for two cases: (a) all galaxies in \TestFields and (b) galaxies in \TestFields with $\text{S/N}_{\text{blend}}\geq5$. A value closer to $0$ for $1-\rm{SSIM}$ on the x-axis represents better performance.}
    \label{fig: Morphology all bands}
\end{figure*} 

Since we observed in Appendix \ref{section: appendix photometry results} that there is a slight dip in the performance of \MADNESS flux reconstructions in the \textit{u}-band, in this Section we want to look into the morphology of the deblended models in different bands. 
Using SSIM as a metric, we show in Figure \ref{fig: Morphology all bands} that \scarlet reconstructs morphology far better than \MADNESS in the \textit{u}-band, although in all the other bands \MADNESS consistently outperforms \scarlet. 

Similar to the Appendix \ref{section: appendix photometry results}, we make the plots for two cases: the first with all the galaxies in \TestFields and the second with only higher S/N galaxies (S/N\textsubscript{blend}$\geq$5). 
In Figure \ref{fig: cumulative_distib_phot_err_masked} we showed that the performance of \MADNESS was comparable to that of \scarlet even in the \textit{u}-band for high S/N galaxies due to the reconstruction term of the MAP optimization.
However, for morphology, in Figure \ref{fig: cosine similarity all bands masked} we do not see such a drastic improvement in the performance of \MADNESS compared to \scarlet in the \textit{u}-band.  
This is because the \scarlet deblender predicts only one morphology in all the bands. Hence, it uses the information from high S/N bands to reconstruct the morphology and this happens to be in sync with the simulations since all bands of a galaxy share the same morphology.
In reality, however, this assumption does not hold. 
So for \MADNESS, we choose to have the flexibility to predict different morphologies in different bands, which makes it difficult to reconstruct the fainter edges in the low S/N bands.

We can try several ways to mitigate this problem. 
First, we can try to add a weighting function to Equation \ref{equation: final deblending solution} to enforce better reconstructions in the \textit{u}-band. 
As an alternative, we can also adopt a non-linear data normalization similar to that of \citet{10.1093/mnras/staa3062} whose authors suppressed the dominance of brighter pixels using the normalization to increase the relative weights of the fainter pixels. 
Finally, we can also try to adopt a \scarlet-like approach to hierarchically predict the morphology and the SED of galaxies but it would constrain one to the unrealistic assumption of the same morphology in all the bands.

\end{appendix}
\end{document}